\documentclass[secnumarabic,amssymb,aps,pra,nobibnotes,showkeys,preprint]{revtex4-1}
\usepackage{docs}
\usepackage[pdftex]{graphicx}
\usepackage{amssymb}
\usepackage{amsmath}
\usepackage{bm}
\usepackage{subfigure}
\usepackage{color}
\expandafter\ifx\csname package@font\endcsname\relax\else
 \expandafter\expandafter
 \expandafter\usepackage
 \expandafter\expandafter
 \expandafter{\csname package@font\endcsname}%
\fi
\DeclareRobustCommand\substyle{\name@idx{document substyle}}%
\DeclareRobustCommand\classoption{\name@idx{document class option}}%
\DeclareRobustCommand\classname{\name@idx{document class}}%
\def\name@idx#1#2{%
 {\ttfamily#2}%
 \index{#2\space#1=\string\ttt{#2}\space#1}\index{#1>#2=\string\ttt{#2}}%
}%

\begin{document}
\title{Two-Dimensional Graphene: Theoretical Study of Multi-photon Non-linear  Absorption Coefficient of a Strong Electromagnetic Wave by Using Quantum Kinetic Equation}

\author{Tran Anh Tuan}
\author{Nguyen Quang Bau}
\email{Corresponding author: nguyenquangbau54@gmail.com, nguyenquangbau@hus.edu.vn}
\author{Nguyen Dinh Nam}
\email{Corresponding author: nguyendinhnam@hus.edu.vn}
\author{Cao Thi Vi Ba}
\email{Corresponding author: caothiviba@hus.edu.vn}
\author{Nguyen Thi Thanh Nhan}
\affiliation{Department of Theoretical Physics, University of Science, Vietnam National University, Hanoi \\ Address: $N^{0}$ 334 Nguyen Trai, Thanh Xuan, Hanoi, Vietnam}
\date{}
%\tableofcontents
\begin{abstract}
Based on the quantum kinetic equation for electrons, we theoretically study the quantum multi-photon non-linear absorption of a strong electromagnetic wave (EMW) in two-dimensional graphene. Two cases of the electron scattering mechanism are considered: Electron-optical phonon scattering and electron-acoustic phonon scattering. The general multi-photon absorption coefficient is presented as a function of the temperature, the external magnetic field, the photon energy and the amplitude of external EMW. These analytical expressions for multi-photon non-linear absorption coefficient (MNAC) are numerically calculated and the results are discussed in both the absence and presence of a magnetic field perpendicular to the graphene sheet. The results show that there is no absorption peak in the absence of the magnetic field, which contrasts with previous results in 2D systems such as quantum wells or superlattices. However, when there is a strong magnetic field along the direction perpendicular to the 2D graphene, absorption spectral lines appear consistent with the magneto-phonon resonance conditions. Our calculations show that the multi-photon absorption's effect is stronger than mono-photon absorption. Besides, the quantum multi-photon non-linear absorption phenomenon has been studied from low to high temperatures. This transcends the limits of the classical Boltzmann kinetic equation which is studied in the high-temperature domain. The computational results show that the dependence of MNAC on the above quantities is consistent with the previous theoretical investigation. Another novel feature of this work is that the general analytic expression for MNAC shows the Half Width at Half Maximum (HWHM) dependence on the magnetic field which is in good agreement with the previous experimental observations. Thus, our estimation might give a critical prediction for future experimental observations in 2D graphene. 
\end{abstract}
\maketitle

\section{Introduction}
In recent times, there has been considerable interest in investigating the analytic behavior of low-dimensional systems, particularly, two-dimensional electron gas (2DEG) systems such as Quantum Wells and Superlattices. Monolayer graphene is one of the most remarkable 2D materials. Graphene is made of carbon atoms in a honeycomb lattice and a unit cell containing two carbon atoms \cite{no1}. More recent research on the electronic structure of graphene show that graphene is called a zero-gap semiconductor \cite{no2} because of the vanishing at energy $E=0$ in the density of states per unit cell, $\rho \left( E \right) = \dfrac{{3\sqrt 3 {a^2}}}{\pi }\dfrac{{\left| E \right|}}{{v_F^2}}$, in which $v_F$ represents Fermi velocity and $a$ is the lattice constant in 2D graphene. Another noteworthy point is that electrons and holes in graphene behave as the massless fermions described by 2D Weyl equations \cite{shen, ando1}. In contrast to traditional 2D semiconductors, where the motion of electrons is restricted to one dimension, they can only flow freely in two dimensions \cite{eps}. 

The above shows that 2D graphene has unique electronic and transport properties. Its properties are the strange interference between traditional semiconductors (zero density of states) and metals (gapless), which make them potential candidates for a wide range of applications in modern electronic technology and one of the most crucial research directions in Condensed Matter Physics as well as Material Science in recent years. 

The Quantum non-linear absorption effect, one of the most significant research topics in low-dimensional electron gas systems has been studied in theoretical and experimental aspects \cite{ni, h, eps1} using different techniques. In traditional low-dimensional systems, because the motion of electrons is restricted to one, two, or three dimensions, the energy spectrum of the electrons confined becomes completely quantized. Thus, this affects many of the optical and electrical properties of the system. The most remarkable thing is that the presence of an EMW propagating in materials increases the probability of scattering between electrons and phonons and many differences compared to the absence of EMW. Furthermore, linear absorption of a weak EMW in conventional semiconductors has been studied using the Kubo-Mori method \cite{bau2}. On the other hand, the non-linear absorption of a strong EMW in low-dimensional electron-gas systems has also been indicated clearly by using the quantum kinetic equation method \cite{phong, van3, bau3}. Especially for graphene, many different research methods have been used to investigate the absorption effect. Bui Dinh Hoi and co-workers \cite{h1} found the explicit expression of optical absorption power in 2D graphene under a perpendicular magnetic field and proved that the half width at half maximum of cyclotron-phonon resonance does not depend on the temperature by using the projection operator technique. Huynh Vinh Phuc and Nguyen Ngoc Hieu \cite{phuc} presented the non-linear optical absorption power in the presence of the perpendicular magnetic field with an electron-optical phonon interaction picture via two-photon absorption process using the perturbation approximation method. When surveying the effect of the EMW on the absorption power for a unit surface, S.V. Kryuchkov \textit{et al.} \cite{kry} pointed out that in the large amplitudes of circularly polarized EMW, the absorption power is proportional to the EMW's amplitude within the complete self-consistent relaxation time approximation. The optical power has been investigated in all these works, taking into account only electron-optical phonon interaction in the high-temperature range. In contrast, the electron-acoustic phonon in the low-temperature range has not been a concern. Moreover, the previous works also did not mention the influence of linearly polarized EMW on the MNAC via the multi-photon absorption processes (MPA). 

Multi-photon absorption (MPA) is a typical optical effect in low-dimensional material research. This phenomenon describes the transition of electrons from the ground state level to the higher state level via the simultaneous absorption of more than one photon that Mayer has predicted since 1931 \cite{may} and Kaiser \textit{et al.} published the first experimental results of the simplest multi-photon absorption effect, the absorption of two photons, in 1961. \cite{kai}. Because of its vital role, MPA has been extensively studied in theoretical and experimental research in most low-dimensional electron systems, in general \cite{morita, goodwin} and two-dimensional electron gas, in particular \cite{bhat1, bhat, lvtung, ando2, ando3}. These results lead to the conclusion of the appearance of Magneto-phonon Resonance, Electron-phonon Resonance, and Cyclotron Resonance at the appropriate values of photon energy. However, the above studies usually only consider the absorption of 1 or 2 photons. The multi-photon absorption processes have not been fully studied. 

In this work, we present our calculation of the non-linear absorption coefficient of an intense EMW in 2D graphene in which both electron-optical phonon scattering and electron-acoustic phonon scattering with the MPA process have been investigated by using the quantum kinetic equation method. This approach allows us to consider the rate of change in unbalanced electron distribution function and, thus shows the electron-phonon interaction picture clearly under the influence of a laser radiation field. This is the reason why this method has been used effectively in the study of EMW absorption effects in bulk semiconductors \cite{eps, eps1}, Doped Semiconductor Superlattices \cite{phong}, Quantum Wells \cite{bau,bau4} and Cylindrical Quantum Wires \cite{phong1}. First, this paper assumes that the electrons are non-degenerate and that the distribution function has the form of the Maxwell-Boltzmann function. Second, electron-phonon interaction is much stronger and worth considering than the others (electron-electron, phonon-phonon, ...), and phonons being the non-polar longitudinal phonons. Third, the electron-acoustic phonon interaction is considered quasi-elastic, and we ignore acoustic phonon energy $\hbar {\omega _{\mathbf{q}}}$ in the  Dirac delta function \cite{van1, van3} since this energy is too small compared to the photon energy of EMW. Our calculations are compared to the recent theoretical results, which show the difference and the novelty of the results. 

We have organized the rest of this paper in the following way. In Sec. 2, we carry out the method for establishing the quantum kinetic equation based on the electronic properties of graphene (the wave function and the eigenenergy). The calculation of the non-linear absorption coefficient is also presented briefly in Sec. 2 and Sec. 3. Then, in Sec. 4, we estimate numerical values, including an analysis of the dependence of MNAC on the temperature, the specific parameters for 2D Graphene, the photon energy $\hbar\Omega$ and the amplitude $E_0$ of EMW. Finally, conclusions are given in Sec. 5.

\section{Multi-photon non-linear absorption coefficient in the case of the absence of an external magnetic field}

In this paper, the 2D graphene sheet is taken along in the $x-y$ plane. The motion of the electrons can be written by the 2D Weyl equation for Dirac massless fermions \cite{ando1}. Hence, the wave function and the eigenenergy of the electron are, respectively, given by \cite{ando4, ando5} 
\begin{align}\label{wf0}
\left| {n,{\bf{k}}} \right\rangle  = \Psi \left( {\bf{r}} \right) = \frac{{\exp \left( {i{\bf{k}} \cdot {\bf{r}}} \right)}}{{L\sqrt 2 }}\left( {\begin{array}{*{20}{c}}
n\\
{{e^{i\theta \left( {\bf{k}} \right)}}}
\end{array}} \right)
\end{align}
\begin{align}
    {\varepsilon _{n,{\mathbf{k}}}}\left({\bf{k}} \right)  = n{\gamma }\left| {\bf{k}} \right|
\end{align}
Where $n =  \pm 1$ denotes the conduction and valence bands and 
\begin{align}
\begin{array}{*{20}{c}}
  {{k_x} = \left| {\bf{k}} \right|\cos \theta \left( {\mathbf{k}} \right)}&{{k_y} = \left| {\bf{k}} \right|\sin \theta \left( {\mathbf{k}} \right)} 
\end{array}
\end{align}
$\mathbf{k}$ and ${L_y} = {L_x} = L$ are the wave vector and the sample normalization length in the directions of the system. The area of the sample is assumed to be $S_0=L_{x}L_{y}=L^2$ and $\gamma  = 6.46{\text{eV}}{\text{.{\AA}}}$ being the band parameter. 

When the EMW is applied to the specimen in the z-direction with the electric field vector ${\mathbf{E}} = \left( {0,{E_0}\sin \Omega t,0} \right)$ ($E_0$ and $\Omega$ are the amplitude and frequency of the EMW), the total Hamiltonian in monolayer graphene in the second quantization representation of our model reads as
\begin{align} \label{ha}
{\mathcal{H}} = {{\mathcal{H}}_{{\rm{electron}}}} + {{\mathcal{H}}_{{\rm{phonon}}}} + {{\mathcal{H}}_{{\rm{electron - electron}}}} + {{\mathcal{H}}_{{\rm{electron - phonon}}}}
\end{align}
where ${{\mathcal{H}}_{{\rm{electron - electron}}}}$ is the electron-electron interaction energy which is assumed to be very weak, and thus we can ignore it. We can write the electronic single-particle Hamiltonian as \cite{ben}
\begin{align}\label{he}
{{\mathcal{H}}_{{\rm{electron}}}} = \sum\limits_{n,{{\bf{k}}_y}} {{\varepsilon _{n,{{\bf{k}}_y}}}\left( {{{\bf{k}}_y} - \frac{e}{{\hbar c}}{\bf{A}}\left( t \right)} \right)a_{n,{{\bf{k}}_y}}^\dag {a_{n,{{\bf{k}}_y}}}} 
\end{align}

And the phononic part of Hamiltonian in Eq. \eqref{ha} \cite{ando2}
\begin{align}\label{hp}
{{\mathcal{H}}_{{\rm{phonon}}}} = \sum\limits_{\bf{q}} {\hbar {\omega _{\bf{q}}}\left( {b_{\bf{q}}^\dag {b_{\bf{q}}} + \frac{1}{2}} \right)} 
\end{align}
Here, $a_{n,{{\mathbf{k}}_y}}^ \dag $ and ${a_{n,{{\mathbf{k}}_y}}}$ $\left( {b_{\mathbf{q}}^ \dag \,{\text{and}}\,{b_{\mathbf{q}}}} \right)$
are the creation and annihilation operators of electron (phonon), respectively; $\hbar {\omega _{\mathbf{q}}}$ is the phonon energy. The vector potential of laser radiation as a strong EMW ${\mathbf{A}}\left( t \right)$ takes the form 
\begin{align}
{\mathbf{A}}\left( t \right) = \frac{c}{\Omega }{{\mathbf{E}}_0}\cos \left( {\Omega t} \right)
\end{align}
 
The interaction part of Hamiltonian in Eq. \eqref{ha} between electron and phonon is written as 
\begin{align}\label{hep}
{{\mathcal{H}}_{{\rm{electron - phonon}}}} = \sum\limits_{n,n'} {\sum\limits_{{{\bf{k}}_y},{{\bf{q}}_y}} {{\rm{C}}\left( {\bf{q}} \right)a_{n',{{\bf{k}}_y} + {{\bf{q}}_y}}^\dag {a_{n,{{\bf{k}}_y}}}\left( {b_{ - {\bf{q}}}^\dag  + {b_{\bf{q}}}} \right)} } 
\end{align}
In Eq. \eqref{hep}, $n$ and $n'$ are the band indices of states $\left| {\bf{k}} \right\rangle $ and $\left| {{\bf{k}} + {\bf{q}}} \right\rangle $, respectively. Here, ${\rm{C}}\left( {\bf{q}} \right)$ is the electron-phonon matrix element which depends on the scattering mechanism.

The general quantum kinetic equation for electron distribution function \cite{bau1,eps, eps1}  
\begin{align}\label{qke}
\frac{{\partial {f_{n,{{\mathbf{k}}_y}}}\left( t \right)}}{{\partial t}} =  - \frac{i}{\hbar }{\left\langle {\left[ {a_{n,{{\mathbf{k}}_y}}^ \dag {a_{n,{{\mathbf{k}}_y}}},{\text{H}}} \right]} \right\rangle _t}
\end{align}
where ${f_{n,{{\mathbf{k}}_y}}}\left( t \right) = {\left\langle {a_{n,{{\mathbf{k}}_y}}^\dag {a_{n,{{\mathbf{k}}_y}}}} \right\rangle _t}$ is the electron distribution function, and  ${\left\langle ...  \right\rangle _t}$ denotes the statistical average value at the moment t. 

To investigate the optical absorption properties in the graphene sheet in the presence of an EMW, we need to have an analytic expression of the non-linear absorption coefficient \cite{eps, bau, bau1}
\begin{align}\label{a}
\alpha  = \frac{{8\pi }}{{c\sqrt {{\kappa _\infty }} E_0^2}}{\left\langle {{{\bf{J}}_y} \cdot {\bf{E}}} \right\rangle _t}
\end{align}
where $\kappa _\infty$ is the high-frequency dielectric constant in graphene and $c$ is the speed of light in a vacuum. 

In Eq. \eqref{a}, ${{\mathbf{J}}_y}\left( t \right)$ is the carrier current density in 2D graphene, which flows in the y-direction because the laser field is polarized in the y-direction. Its expression is given as \cite{bau}
\begin{align}\label{j}
{{\mathbf{J}}_y}\left( t \right) = \frac{{e\hbar }}{{{m_e}}}\sum\limits_{n,{{\mathbf{k}}_y}} {\left( {{{\mathbf{k}}_y} - \frac{e}{{\hbar c}}{\mathbf{A}}\left( t \right)} \right){f_{n,{{\mathbf{k}}_y}}}\left( t \right)} 
\end{align}
Here, $e$ and $m_e=0.012m_0$ \cite{me} are the charges and the effective mass of electrons in graphene, $m_0$ is the electron rest mass. 

Starting from Hamiltonian ${\text{H}}$ in Eq. \eqref{ha} and doing operator algebraic calculations, then using the first-order tautology approximation method to solve this equation as in \cite{bau}, we obtain the time evolution of the distribution function for electrons in graphene as 

\begin{align}\label{evo}
\begin{split}
&\frac{{\partial {f_{n,{{\bf{k}}_y}}}\left( t \right)}}{{\partial t}} = \frac{1}{{{\hbar ^2}}}\sum\limits_{n',{\bf{q}}} {{{\left| {{\rm{M}}\left( {\bf{q}} \right)} \right|}^2}\sum\limits_{\ell ,s} {{J_\ell }\left[ {\frac{{e{E_0}\gamma }}{{{\hbar ^2}{\Omega ^2}}} \left( {n' - n} \right)} \right]{J_s}\left[ {\frac{{e{E_0}\gamma }}{{{\hbar ^2}{\Omega ^2}}} \left( {n' - n} \right)} \right]{\rm{exp}}\left[ {{\rm{i}}\left( {\ell  - s} \right)\Omega {\rm{t}}} \right]} } \\
& \times \int\limits_0^{ + \infty } {d{t_1}\left\{ {\left[ {{f_{n',{{\bf{k}}_y} + {{\bf{q}}_y}}}\left( {{t_1}} \right)\left( {{N_{\bf{q}}}\left( {{t_1}} \right) + 1} \right) - {f_{n,{{\bf{k}}_y}}}\left( {{t_1}} \right){N_{\bf{q}}}\left( {{t_1}} \right)} \right]} \right.{\rm{exp}}\left[ {\frac{{\rm{i}}}{\hbar }\left( {{\varepsilon _{{\rm{n'}},{{\bf{k}}_{\rm{y}}}{\rm{ + }}{{\bf{q}}_{\rm{y}}}}} - {\varepsilon _{{\rm{n}},{{\bf{k}}_{\rm{y}}}}} - \hbar {\omega _{\bf{q}}} - \ell \hbar \Omega } \right)\left( {t - {t_1}} \right)} \right]} \\
& + \left[ {{f_{n',{{\bf{k}}_y} + {{\bf{q}}_y}}}\left( {{t_1}} \right){N_{\bf{q}}}\left( {{t_1}} \right) - {f_{n,{{\bf{k}}_y}}}\left( {{t_1}} \right)\left( {{N_{\bf{q}}}\left( {{t_1}} \right) + 1} \right)} \right]{\rm{exp}}\left[ {\frac{i}{\hbar }\left( {{\varepsilon _{n',{k_y} + {q_y}}} - {\varepsilon _{n,{k_y}}} + \hbar {\omega _q} - \ell \hbar \Omega } \right)\left( {t - {t_1}} \right)} \right]\\
& - \left[ {{f_{n',{{\bf{k}}_y}}}\left( {{t_1}} \right)\left( {{N_{\bf{q}}}\left( {{t_1}} \right) + 1} \right) - {f_{n,{{\bf{k}}_y} - {{\bf{q}}_y}}}\left( {{t_1}} \right){N_{\bf{q}}}\left( {{t_1}} \right)} \right]{\rm{exp}}\left[ {\frac{{\rm{i}}}{\hbar }\left( {{\varepsilon _{{\rm{n}},{{\bf{k}}_{\rm{y}}}}} - {\varepsilon _{{\rm{n'}},{{\bf{k}}_{\rm{y}}}{\rm{ - }}{{\bf{q}}_{\rm{y}}}}} - \hbar {\omega _{\bf{q}}} - \ell \hbar \Omega } \right)\left( {t - {t_1}} \right)} \right]\\
& - \left[ {{f_{n',{{\bf{k}}_y}}}\left( {{t_1}} \right){N_{\bf{q}}}\left( {{t_1}} \right) - {f_{n,{{\bf{k}}_y} - {{\bf{q}}_y}}}\left( {{t_1}} \right)\left( {{N_{\bf{q}}}\left( {{t_1}} \right) + 1} \right)} \right]\left. {{\rm{exp}}\left[ {\frac{{\rm{i}}}{\hbar }\left( {{\varepsilon _{{\rm{n}},{{\bf{k}}_{\rm{y}}}}} - {\varepsilon _{{\rm{n'}},{{\bf{k}}_{\rm{y}}}{\rm{ - }}{{\bf{q}}_{\rm{y}}}}}{\rm{ + }}\hbar {\omega _{\bf{q}}} - \ell \hbar \Omega } \right)\left( {t - {t_1}} \right)} \right]} \right\}
\end{split}
\end{align}
where $J_{s}(x)$ is the $s$th-order Bessel function of the argument $x$. ${{\bar f}_{n,{{\bf{k}}_y}}}$ is the equilibrium distribution function for electrons and be assumed to obey the Maxwell-Boltzmann distribution function \cite{bau, h1}, in which ${{\varepsilon _F}}$, $n_0$ is the Fermi energy and the electron density in graphene. $k_B$ is the Boltzmann constant, and T is the absolute temperature of the system. ${\overline N _{\bf{q}}}$ is the equilibrium distribution function for phonons, which is given by the Bose-Einstein distribution function ${\overline N _{\bf{q}}} = {\left[ {\exp \left( {{{\hbar {\omega _{\bf{q}}}} \mathord{\left/
 {\vphantom {{\hbar {\omega _{\bf{q}}}} {{k_B}T}}} \right.
 \kern-\nulldelimiterspace} {{k_B}T}}} \right) - 1} \right]^{ - 1}}$.

Substituting ${{f_{n,{{\bf{k}}_y}}}\left( t \right)}$ in Eq. \eqref{evo} into the expression for ${{\bf{J}}_y}\left( t \right)$ in Eq. \eqref{j} and inserting in Eq. \eqref{a}, we can find the general multi-photon  non-linear absorption coefficient of the electromagnetic wave owing to the absorption of $\ell$ photons  
\begin{align}\label{abs}
\alpha  = \frac{{8{\pi ^2}e}}{{c\sqrt {{\kappa _\infty }} {E_0}{m_e}\Omega }}\sum\limits_{n,n'} {\sum\limits_{{{\bf{k}}_y},{\bf{q}}} {\frac{{{{\left| {C\left( {\bf{q}} \right)} \right|}^2}}}{{\hbar {\omega _{\bf{q}}}}}q{\overline N _{\bf{q}}}{{\bar f}_{n,{{\bf{k}}_y}}}\sum\limits_\ell  {{{\left( {\frac{{e{E_0}\gamma }}{{{\hbar ^2}{\Omega ^2}}}} \right)}^{2\ell  - 1}}\frac{{\delta \left( {{\varepsilon _{n',{{\bf{k}}_y} + {{\bf{q}}_y}}} - {\varepsilon _{n,{{\bf{k}}_y}}} + \hbar {\omega _{\bf{q}}} - \ell \hbar \Omega } \right)}}{{\ell{{\left[ {\Gamma \left( \ell  \right)} \right]}^2}}}} } } 
\end{align}
Here, $\delta \left( x \right)$ is the Dirac delta function, and ${\Gamma \left( \ell  \right)}$ is the Gamma function for positive integer argument $\ell$. 

In the next step, we will consider two scattering mechanisms of electrons and phonons to elucidate the physical properties involved. 
\subsection{Electron-optical phonon scattering}
At the high temperature $\left( {T > 50{\rm{K}}} \right)$, we assume that the dispersion of phonon follows $\hbar {\omega _{\bf{q}}} \approx \hbar {\omega _0}$ and the matrix element of Hamiltonian for electron-optical phonon scattering is determined by 
\begin{align}\label{mop}
{\left| {{\rm{C}}\left( {\bf{q}} \right)} \right|^2} = \frac{{{\hbar ^2}D_{op}^2}}{{2\rho {L^2}\left( {\hbar {\omega _0}} \right)}}
\end{align}
with $D_{op}$ is the deformation potential coupling constant, and $\rho$ is the graphene mass density.
We use the integral forms of the algebraic summations of ${{{\bf{k}}_y}}$ and ${{{\bf{q}}_y}}$ and the property of the Dirac delta  function as in previous studies \cite{bau, phong, phong1} so that we find out the expression for the non-linear optical absorption coefficient in the case of electron-optical phonon scattering in the following equation
 
\begin{align}\label{aop}
{\alpha _{op}} = \frac{{8{n_0}{{\left( {{k_B}T} \right)}^2}\hbar \Omega D_{op}^2{{\overline N }_{\bf{q}}}}}{{c\sqrt {{\kappa _\infty }} E_0^2{m_e}\rho {\gamma ^5}{\omega _0}}}\sum\limits_\ell  {\frac{\ell }{{{{\left( {\ell !} \right)}^2}}}{{\left( {\frac{{e{E_0}\gamma }}{{{\hbar ^2}{\Omega ^2}}}} \right)}^{2\ell }}\left( {\ell \hbar \Omega  - \hbar {\omega _0}} \right)\exp \frac{{\ell \hbar \Omega  - \hbar {\omega _0}}}{{2{k_B}T}}} 
\end{align}

It can be seen that MNAC depends on the absolute temperature T of the system, the characteristic parameters in a graphene sample, the photon energy, and the intensity of the laser field (EMW). 
\subsection{Electron-acoustic phonon scattering}
For this scattering mechanism, we need to do the same calculation steps as in the case of electron-optical phonon scattering and use Eq. \eqref{abs}. However, there is some variation in the matrix element formula as follows \cite{kuba, bhat, vasko}  
\begin{align}\label{mac}
{\left| {{\rm{C}}\left( {\bf{q}} \right)} \right|^2} = \frac{{\hbar D_{ac}^2q}}{{2\rho {v_s}{L^2}}}
\end{align}
here, $v_s$ and $D_{ac}$ are the sound velocity in graphene and the magnitude of the deformation potential energy, respectively. Moreover, we assume that the scattering between electron and acoustic phonon is quasi-elastic, hence acoustic phonon energy $\hbar {\omega _{\bf{q}}} = \hbar {v_s}q$ is ignored in Dirac delta function \cite{van1, van3}. Considering the case of a non-degenerate phonon system, the equilibrium distribution function of the acoustic phonon has the form \cite{hwang} ${\overline N _{\bf{q}}} \approx {{{k_B}T} \mathord{\left/
 {\vphantom {{{k_B}T} {\hbar {\omega _{\bf{q}}}}}} \right.
 \kern-\nulldelimiterspace} {\hbar {\omega _{\bf{q}}}}}$ 

Following, by using some same calculations in above case and the properties of Bessel functions, we obtain the general MNAC for the case of electron-acoustic phonon interaction (without magnetic  field) 
\begin{align}
{\alpha _{ac}} = \frac{{8{n_0}{{\left( {{k_B}T} \right)}^3}\hbar \Omega D_{ac}^2}}{{c\sqrt {{\kappa _\infty }} E_0^2{m_e}\rho {\gamma ^5}v_s^2}}\sum\limits_\ell  {\frac{\ell }{{{{\left( {\ell !} \right)}^2}}}{{\left( {\frac{{e{E_0}\gamma }}{{{\hbar ^2}{\Omega ^2}}}} \right)}^{2\ell }}\left( {\ell \hbar \Omega } \right)\exp \frac{{\ell \hbar \Omega }}{{2{k_B}T}}} 
\end{align}

Thus, the quantum optical absorption effect in graphene without a magnetic field has been studied from low to high temperatures with two scattering mechanisms. The expressions for the MNAC in these two cases are very complex because of the differences in the energy spectrum and the wave function of the moving electrons in the graphene lattice. In the next section, we will study the effect of a strong magnetic field on the MNAC of EMW in graphene.

\section{Multi-photon non-linear absorption coefficient in the case of the presence of an external magnetic field}
We considered a graphite sheet where electrons move freely in the $(x,y)$ plane. The system was subjected to a magnetic field $\mathbf{B}=(0,0,B)$ with a vector potential $\mathbf{A} ' = \left( {0,Bx,0} \right)$. Energy levels in strong magnetic field B are quantized into discrete energy levels, called Landau levels or Landau orbits. The wave function and the corresponding energy are written as  \cite{ando2,yang} 
 \begin{align}\label{wf}
 \Psi \left( {\mathbf{r}} \right) \equiv \left| {n,{{\mathbf{k}}_y}} \right\rangle  = \frac{{{C_n}}}{{\sqrt L }}\exp \left( { - i\frac{{Xy}}{{{l_{B}^2}}}} \right)\left[ {\begin{array}{*{20}{c}}
  {{S_n}{\Phi _{\left| n \right| - 1}}\left( {x - X} \right)} \\ 
  {{\Phi _{\left| n \right|}}\left( {x - X} \right)} 
\end{array}} \right] = \frac{{{C_n}}}{{\sqrt {{L_y}} }}{e^{i{k_y}y}}\left[ {\begin{array}{*{20}{c}}
  {{S_n}{\Phi _{\left| n \right| - 1}}\left( {x - X} \right)} \\ 
  {{\Phi _{\left| n \right|}}\left( {x - X} \right)} 
\end{array}} \right]
 \end{align}
\begin{align}\label{el}
{\varepsilon _n} = {S_n}\hbar {\omega _B}\sqrt {\left| n \right|} 
\end{align}
with $S_n$ is the sign function defined by $S_n=-1$ for an
electron when $n<0$, $S_n= 1$ for a hole when $n>0$, and $S_n
=0$ for $n=0$; $C_n^2 = {{\left( {1 + {\delta _{n,0}}} \right)} \mathord{\left/
 {\vphantom {{\left( {1 + {\delta _{n,0}}} \right)} 2}} \right.
 \kern-\nulldelimiterspace} 2}$ where $\delta_{ij}$ is the Kronecker Delta, $\delta_{ij} = 1$ when $i=j$ and $\delta_{ij} =0$ when $i \ne j$. In Eq. \eqref{wf} $\Phi _{\left| n \right|}\left( x \right)$ is the normalized harmonic oscillator function given by 
\begin{align}
{\Phi _{\left| n \right|}}\left( x \right) = \frac{{{i^{\left| n \right|}}}}{{\sqrt {{2^{\left| n \right|}}\left| n \right|!\sqrt \pi  l_B} }}\exp \left[ { - \frac{1}{2}{{\left( {\frac{x}{l_B}} \right)}^2}} \right]{H_{\left| n \right|}}\left( {\frac{x}{l_B}} \right)
\end{align}
Where $n = 0, \pm 1, \pm 2,...$ being the Landau indices. $X$ being the coordinate of the center of the carrier orbit $X = {k_y}{l_B^2}$ with $l_B = \sqrt {\hbar /eB} $ is the radius of the ground state electron orbit in the $\left( {x,y} \right)$ plane or the magnetic length. ${\mathcal{H}_{\left| n \right|}}\left( {{x \mathord{\left/
 {\vphantom {x l}} \right.
 \kern-\nulldelimiterspace} l_B}} \right)$ is the n-th order Hermite polynomial and $\hbar {\omega _B} = {{\sqrt 2 \gamma } \mathord{\left/
 {\vphantom {{\sqrt 2 \gamma } l_B}} \right.
 \kern-\nulldelimiterspace} l_B}$ is the effective magnetic energy \cite{ando2,ando3}
with $\gamma  = \left( {{{\sqrt 3 } \mathord{\left/
 {\vphantom {{\sqrt 3 } 2}} \right.
 \kern-\nulldelimiterspace} 2}} \right)a{\gamma _0}= 6,46{\text{eV}}{\text{.}}\mathop {\text{A}}\limits^{\text{o}}$  is the band parameter, $\gamma _0=3,03$eV is the resonance integral between nearest neighbor carbon atoms, $a=0,246$ nm is the lattice constant. 

When a strong EMW is applied to the system with the electric field vector ${\mathbf{E}} = \left( {0,{E_0}\sin \Omega t,0} \right)$ ($E_0$ and $\Omega$ are the amplitude and frequency, respectively), the Hamiltonian of the electron-phonon system in the second quantization representation can be written as \cite{ando2, phuc}
\begin{align}\label{ha1}
\mathcal{H} = \sum\limits_{n,{{\mathbf{k}}_y}} {{\varepsilon _n}\left( {{{\mathbf{k}}_y} - \frac{e}{{\hbar c}}{\mathbf{A}}\left( t \right)} \right)a_{n,{{\mathbf{k}}_y}}^\dag {a_{n,{{\mathbf{k}}_y}}}} + \sum\limits_{\mathbf{q}} {\hbar {\omega _{\mathbf{q}}}\left( {b_{\mathbf{q}}^\dag {b_{\mathbf{q}}} + \frac{1}{2}} \right)} 
+ \sum\limits_{n,n'} {\sum\limits_{{{\mathbf{k}}_y},{\mathbf{q}}} {{\rm{M}_{n.n'}}\left( {\mathbf{q}} \right)a_{n',{{\mathbf{k}}_y} + {{\mathbf{q}}_y}}^\dag {a_{n,{{\mathbf{k}}_y}}}\left( {b_{ - {\mathbf{q}}}^\dag  + {b_{\mathbf{q}}}} \right)} } 
\end{align}
For the calculation of this article, we need $\rm{M}_{n.n'}(\mathbf{q})$ (the matrix factor) as follows \cite{phuc} 
 \begin{align}
 {\left| {{\rm{M}_{n,n'}}\left( {\mathbf{q}} \right)} \right|^2} = {\left| {\rm{C}\left( {\mathbf{q}} \right)} \right|^2}{\left| {{\rm{J}_{n,n'}}\left( {\mathbf{q}} \right)} \right|^2}
 \end{align}
with $\rm{C}\left( {\mathbf{q}} \right)$ is the electron-phonon interaction constant which depends on the scattering mechanism according to Eq. \eqref{mop} and Eq. \eqref{mac}; ${{\rm{J}_{n,n'}}\left( {\mathbf{q}} \right)}$ given by 
 \begin{align}
 {\left| {{\rm{J}_{n,n'}}\left( {\mathbf{q}} \right)} \right|^2} = C_n^2C_{n'}^2\frac{{m!}}{{\left( {m + j} \right)!}}{e^{ - u}}{u^j}{\left[ {L_m^j\left( u \right) + {S_n}{S_{n'}}\sqrt {\frac{{m + j}}{m}} L_{m - 1}^j\left( u \right)} \right]^2}
 \end{align}
with ${L_m^j\left( x \right)}$ is the associated Laguerre polynomial, $u = {l_B^2}{q^2}/2,{q^2} = q_x^2 + q_y^2,m = \min \left( {\left| n \right|,\left| {n'} \right|} \right),j = \left| {\left| {n'} \right| - \left| n \right|} \right|$

Solving the quantum kinetic equation, which is established for electrons in graphene for the case of an external magnetic field, by using a method similar to that used to solve Eq. \eqref{evo}, we obtain the $\ell$-photon absorption coefficient 
\begin{align}\label{abtq}
    \alpha  = \frac{{8{\pi ^2}\Omega }}{{c\sqrt {{\kappa _\infty }} E_0^2}}\sum\limits_{{\bf{q}},n',n} {{{\left| {{{\rm{M}}_{n,n'}}\left( {\bf{q}} \right)} \right|}^2}{{\overline N }_{\bf{q}}}\sum\limits_\ell  {\ell J_\ell ^2\left( {\frac{{e{E_0}q}}{{2{m_e}{\Omega ^2}}}} \right)\sum\limits_{{{\bf{k}}_y}} {\overline f \left( {{\varepsilon _n}} \right)\delta \left( {{\varepsilon _{n'}} - {\varepsilon _n} + \hbar {\omega _{\bf{q}}} - \ell \hbar \Omega } \right)} } } 
\end{align}
Here, we assume that the distribution function of electrons in thermal equilibrium is the Fermi-Dirac distribution $\overline f \left( {{\varepsilon _n}} \right) = {\left\{ {1 + \exp \left[ {{{\left( {{\varepsilon _n} - {\varepsilon _F}} \right)} \mathord{\left/
 {\vphantom {{\left( {{\varepsilon _n} - {\varepsilon _F}} \right)} {\left( {{k_B}T} \right)}}} \right.
 \kern-\nulldelimiterspace} {\left( {{k_B}T} \right)}}} \right]} \right\}^{ - 1}}$, in which, $\varepsilon_{F}$ is Fermi energy. 
\subsection{Electron-optical phonon scattering}
Transforming the summations over ${\bf{q}}$ and ${\bf{k}}_y$ to integrals as follows \cite{h1} and using the approximate expression of the Bessel function, we obtain the absorption coefficient for the case of electron-optical phonon interaction 

\begin{align}\label{aotq}
   \alpha _{op}^B = \frac{{4\pi D_{op}^2\hbar \Omega {{\overline N }_{\bf{q}}}}}{{c\sqrt {{\kappa _\infty }} E_0^2\rho l_B^2{\omega _0}}}\sum\limits_\ell  {{{\left( {\frac{{e{E_0}}}{{2{m_e}{\Omega ^2}}}} \right)}^{2\ell }}\frac{1}{{\ell {{\left[ {\Gamma \left( \ell  \right)} \right]}^2}}}\sum\limits_{n,n'} {\bar f\left( {{\varepsilon _n}} \right)\delta \left( {{\varepsilon _{n'}} - {\varepsilon _n} + \hbar {\omega _0} - \ell \hbar \Omega } \right)\int\limits_0^{ + \infty } {{q^{2\ell  + 1}}{{\left| {{{\rm{J}}_{n,n'}}\left( u \right)} \right|}^2}dq} } } 
\end{align}

Using the integrals in the appendix and analytic transformations, we obtain an explicit expression of the optical absorption coefficient in the case of electron-optical phonon scattering
\begin{align}\label{aob}
\alpha _{op}^B = \frac{{4\pi {n_0}D_{op}^2\hbar \Omega {{\overline N }_{\bf{q}}}}}{{c\sqrt {{\chi _\infty }} E_0^2\rho l_B^2{\omega _0}}}\sum\limits_{n',n} {\bar f\left( {{\varepsilon _n}} \right)\sum\limits_\ell  {{A_\ell }\delta \left( {{\varepsilon _{n'}} - {\varepsilon _n} + \hbar {\omega _0} - \ell \hbar \Omega } \right)} } 
\end{align}

$A_\ell$ is the dimensionless parameter characterizing $\ell$-photon absorption process. 

For mono-photon absorption (1PA)
\begin{align}
    {A_1} = {\left( {\frac{{e{E_0}}}{{4{m_e}{\Omega ^2}}}} \right)^2}\frac{{2C_n^2C_{n'}^2}}{{l_B^4}}\left[ {2m + j + 1 - 2{S_n}{S_{n'}}\sqrt {m\left( {m + j} \right)}  + S_n^2S_{n'}^2\left( {2m + j - 1} \right)} \right]
\end{align}

For two-photon absorption (2PA)
\begin{align}
\begin{split}
{A_2} &= {\left( {\frac{{e{E_0}}}{{4{m_e}{\Omega ^2}}}} \right)^4}\frac{{2C_n^2C_{n'}^2}}{{l_B^6}}\left\{ {2 + 6m\left( {m + 1} \right) + j\left[ {j + 3\left( {2m + 1} \right)} \right] - 4{S_n}{S_{n'}}\left( {2m + j} \right)\sqrt {m\left( {m + j} \right)}  + } \right.\\
&\left. { + S_n^2S_{n'}^2\left\{ {2 + 6m\left( {m - 1} \right) + j\left[ {j + 3\left( {2m - 1} \right)} \right]} \right\}} \right\}    
\end{split}
\end{align}

For three-photon absorption (3PA) 
\begin{align}
\begin{split}
{A_3} &= {\left( {\frac{{e{E_0}}}{{4{m_e}{\Omega ^2}}}} \right)^6}\frac{{2C_n^2C_{n'}^2}}{{3l_B^8}}\left\{{\left( {2m + j + 3} \right)\left\{ {2 + 6m\left( {m + 1} \right) + j\left[ {j + 3\left( {2m + 1} \right)} \right]} \right\} + 4m\left( {2m + j} \right)\left( {m + j} \right) + } \right.\\
 &+ S_n^2S_{n'}^2\left\{ {\left( {2m + j + 1} \right)\left\{ {2 + 6m\left( {m - 1} \right) + j\left[ {j + 3\left( {2m - 1} \right)} \right]} \right\} + 4\left( {m - 1} \right)\left( {2m + j - 2} \right)\left( {m + j} \right)} \right\} - \\
&\left. {- 6{S_n}{S_{n'}}\left( {5{m^2} + 5mj + {j^2} + 1} \right)\sqrt {m\left( {m + j} \right)} } \right\}   
\end{split}
\end{align}

Expression (25) shows the highly complex dependence of the optical absorption coefficient on the parameters related to the external field. This is due to the presence of the factor $\rm{J}_{n',n}(u)$, which represents the influence of the magnetic field. 

\subsection{Electron-acoustic phonon scattering}
For this case, the electron-acoustic phonon interaction constant is given by Eq. \eqref{mac}. By the same calculation as above, we can establish the analytic expression of the general multi-photon  non-linear absorption coefficient in the acoustic phonon-electron scattering mechanism 
\begin{align}\label{acb}
    \alpha _{ac}^B = \frac{{4\pi {n_0}{k_B}TD_{ac}^2\Omega }}{{c\sqrt {{\chi _\infty }} E_0^2\rho l_B^2v_s^2}}\sum\limits_{n',n} {\bar f\left( {{\varepsilon _n}} \right)\sum\limits_\ell  {{A_\ell }\delta \left( {{\varepsilon _{n'}} - {\varepsilon _n} - \ell \hbar \Omega } \right)} } 
\end{align}
A remarkable point in Eq. \eqref{aob} and Eq. \eqref{acb} is that the Dirac delta functions will be divergent if their argument approaches zero. To eliminate divergence, we replace the delta functions with the Lorentzian functions as follows \cite{van3, h1} 
\begin{align}
   \delta \left( \varepsilon  \right) = \frac{1}{\pi }\frac{\Gamma }{{{\varepsilon ^2} + {\Gamma ^2}}}
\end{align}
with $\Gamma$ is the dimensionless parameter characterizing the scattering strength, its expression is given \cite{phuc, van3, h1}
\begin{align} 
    {\Gamma ^2} = \sum\limits_{\bf{q}} {\left( {\overline {{N_{\bf{q}}}}  + \frac{1}{2} \pm \frac{1}{2}} \right){{\left| {{\rm{M}}\left( {\bf{q}} \right)} \right|}^2}} 
\end{align}

The analytic expressions of the general MNAC in the presence of a magnetic field are incredibly complex. In the next section, we will give more insight into this difference through numerical computation and graphing with the help of computer programs and numerical methods.

\section{Numerical results and discussions}
In this section, we detail the numerical evaluation of MNAC in both the absence and presence of strong magnetic fields for two electron-phonon scattering mechanisms, from low-temperature to high-temperature domain. The parameters used in computational calculations are as follows: \cite{ando2,ji,h1,phuc,kuba,kry,nose}  $\gamma = 6.46{\text{eV}}{\text{.{\AA}}}, \rho = 7.{7\times 10^{ - 8}}$ $\dfrac{\rm{g}}{{\rm{cm}}^{\rm{2}}}, n_0 = 5 \times 10^{15} \rm{m}^{-2}, {D_{op}} = 1.4 \times {10^9}{\text{eV/cm}}, {D_{ac}} = 19{\text{eV}},{v_s} = 2 \times {10^6}{\text{cm/s}}, \hbar {\omega _0} = 162{\text{meV}},  \kappa_\infty = 4$. The value of the Fermi energy level in the Fermi - Dirac distribution function can be approximated between the Landau levels n = 0 and n = 1 for electrons \cite{ji}. In other words, ${\varepsilon_{F}}$ = ${{\hbar {\omega _B}} \mathord{\left/
 {\vphantom {{\hbar {\omega _B}} 2}} \right.
 \kern-\nulldelimiterspace} 2}$, with $\hbar \omega_B$ is the effective magnetic energy from Eq. \eqref{el}. This paper considers the Landau levels from $-4$ to $4$. 

\subsection{In the Absence of an External Magnetic Field}

\begin{figure}
\centering
\subfigure[][Electron – Optical phonon interaction: $\hbar \Omega  = 450{\rm{meV}}$]{\label{aot0b}\includegraphics[width=0.45\linewidth]{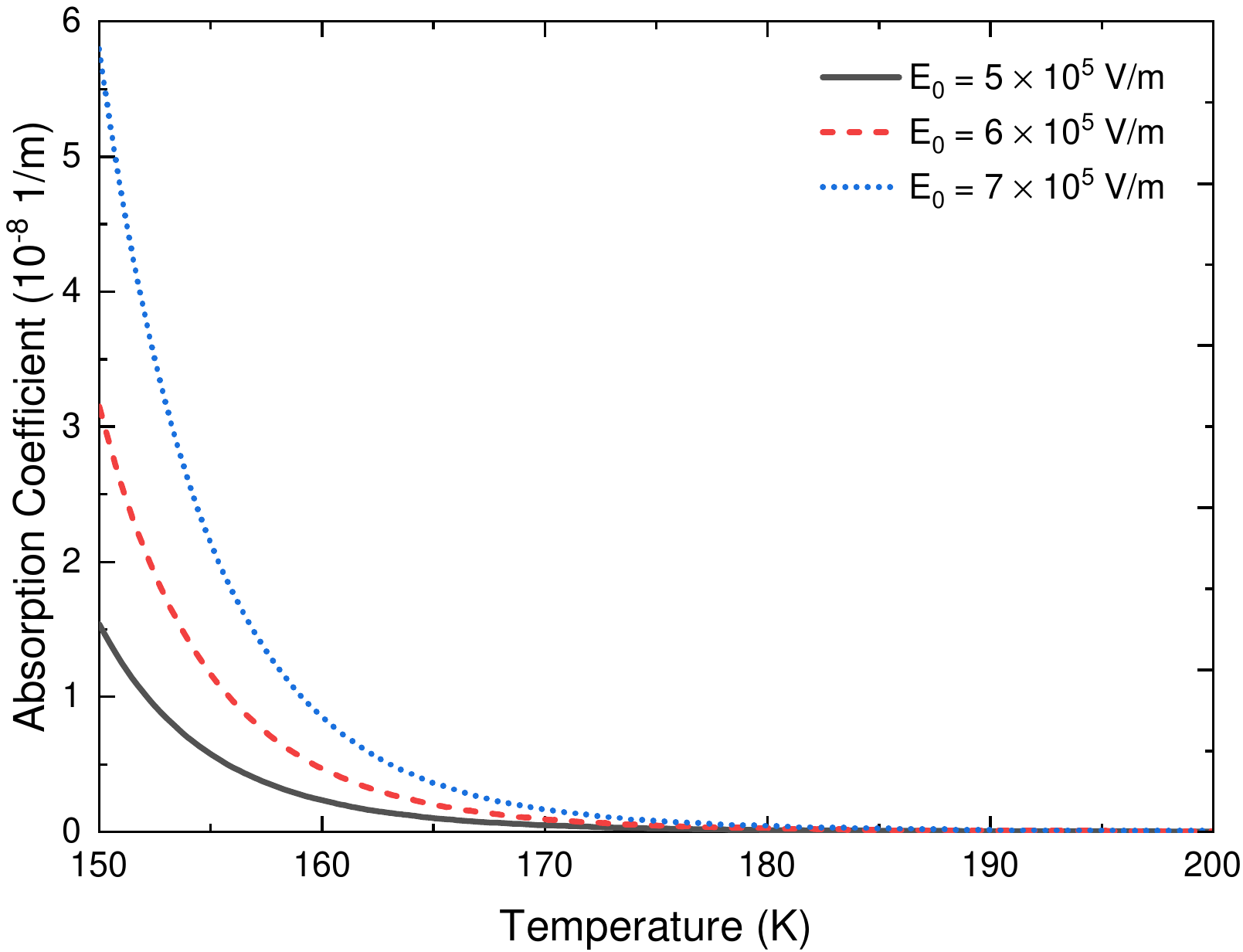}}
\hspace{1pt}
\subfigure[][Electron – Acoustic phonon interaction: $\hbar \Omega  = 50{\rm{meV}}$]{\label{aact0b}\includegraphics[width=0.45\linewidth]{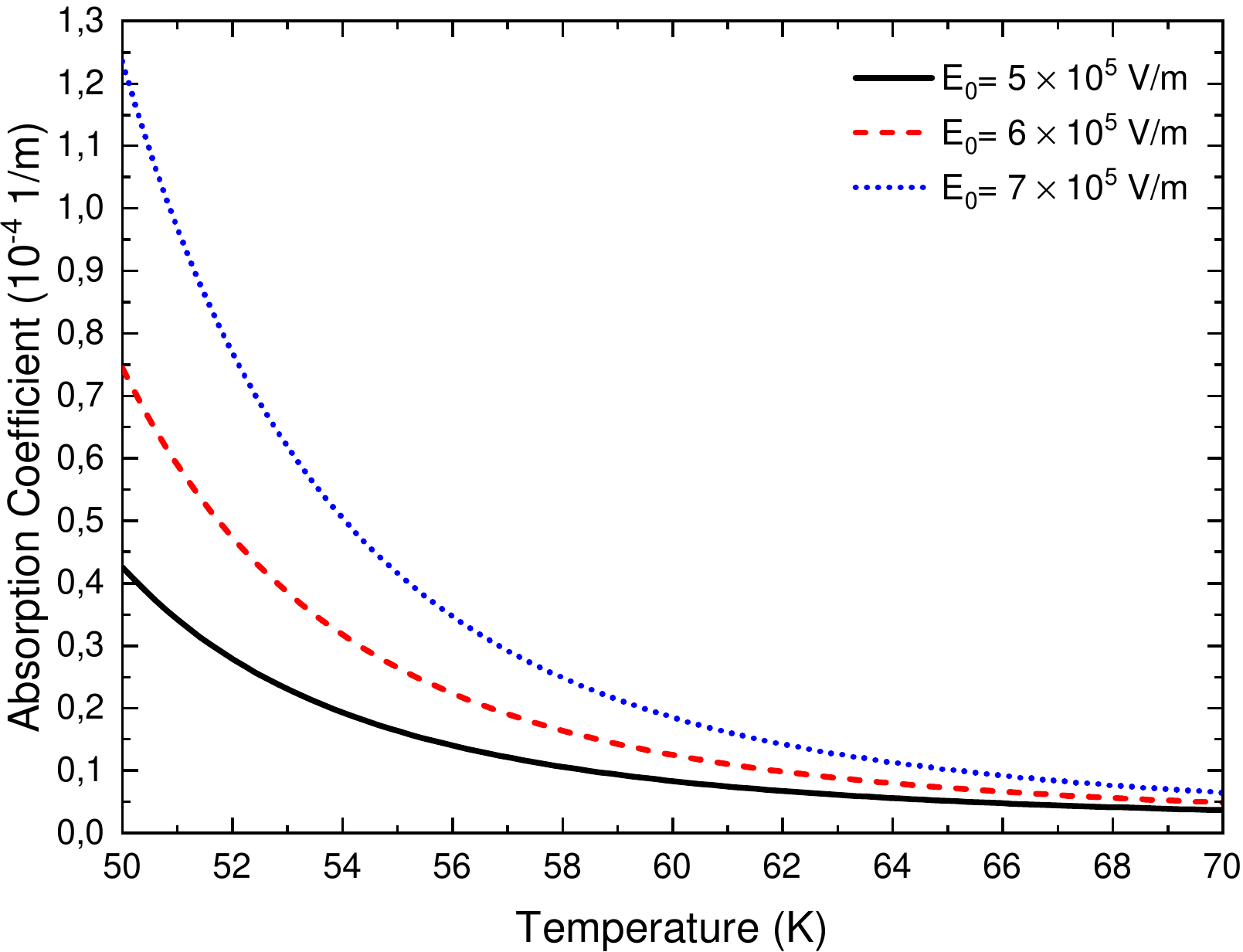}}
\caption{(Color online) The dependence of $\alpha$ on the temperature of the system T}
\label{aT0B}
\end{figure}

The effect of temperature on the absorption coefficient due to electron-phonon scattering is illustrated in Fig. \ref{aT0B}. As can be seen from this figure, the absorption coefficient depends strongly and non-linearly on the system's temperature. We can see that the absorption coefficient reaches the saturated value in the high-temperature regime and decreases quickly when the temperature is high. This trend could be explained that the probability of electron-phonon scattering increases with increasing temperature in two cases, leading to the decrease of the conductivity and then a decrease in the absorption coefficient according to the equation below \cite{bau2} $\label{al}
{\alpha_{xx}}\left( \Omega  \right) = \frac{{4\pi }}{{c{N^*}}}{\mathop{\rm Re}\nolimits} {\sigma _{xx}}\left( \Omega  \right)$. Here, ${N^*}$ is the refraction index and ``Re" denotes ``the real part of". This is a good agreement with previous results using a similar method in a two-dimensional system \cite{bau} and the Boltzmann equation in graphene \cite{kry1}. In addition, we survey the influence of EMW on the absorption coefficient in three different single values of the amplitude of the electric field. It can be seen that when the amplitude of the electric field increases, the absorption coefficient value moves up with the same tendency.

\begin{figure}
\centering
\subfigure[][Electron – Optical phonon interaction: $\hbar \Omega  = 450{\rm{meV}}$\label{aoe00b}]
  {\includegraphics[width=0.45\linewidth]{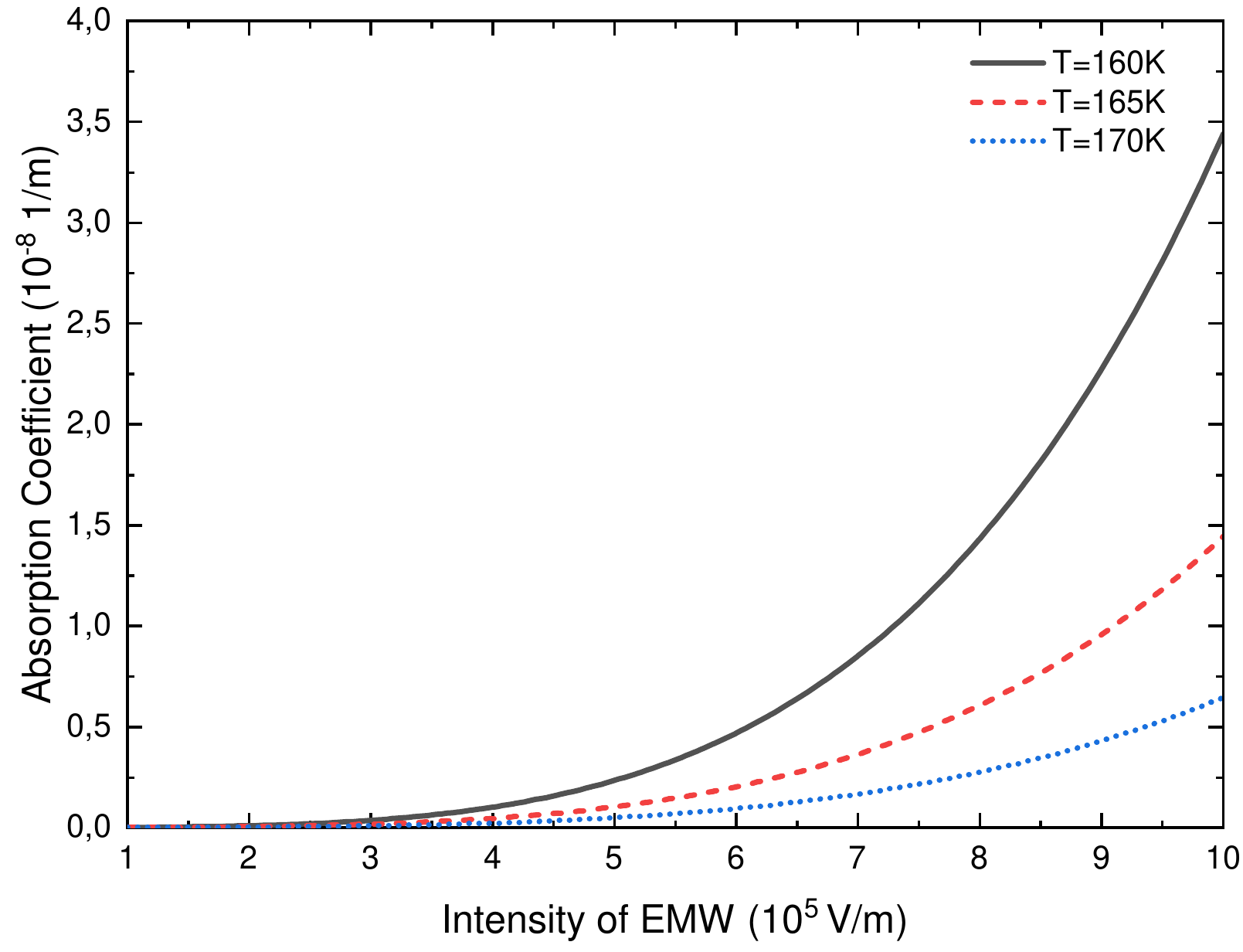}}
\subfigure[][Electron – Acoustic phonon interaction: $\hbar \Omega  = 50{\rm{meV}}$\label{aace00b}]
  {\includegraphics[width=0.45\linewidth]{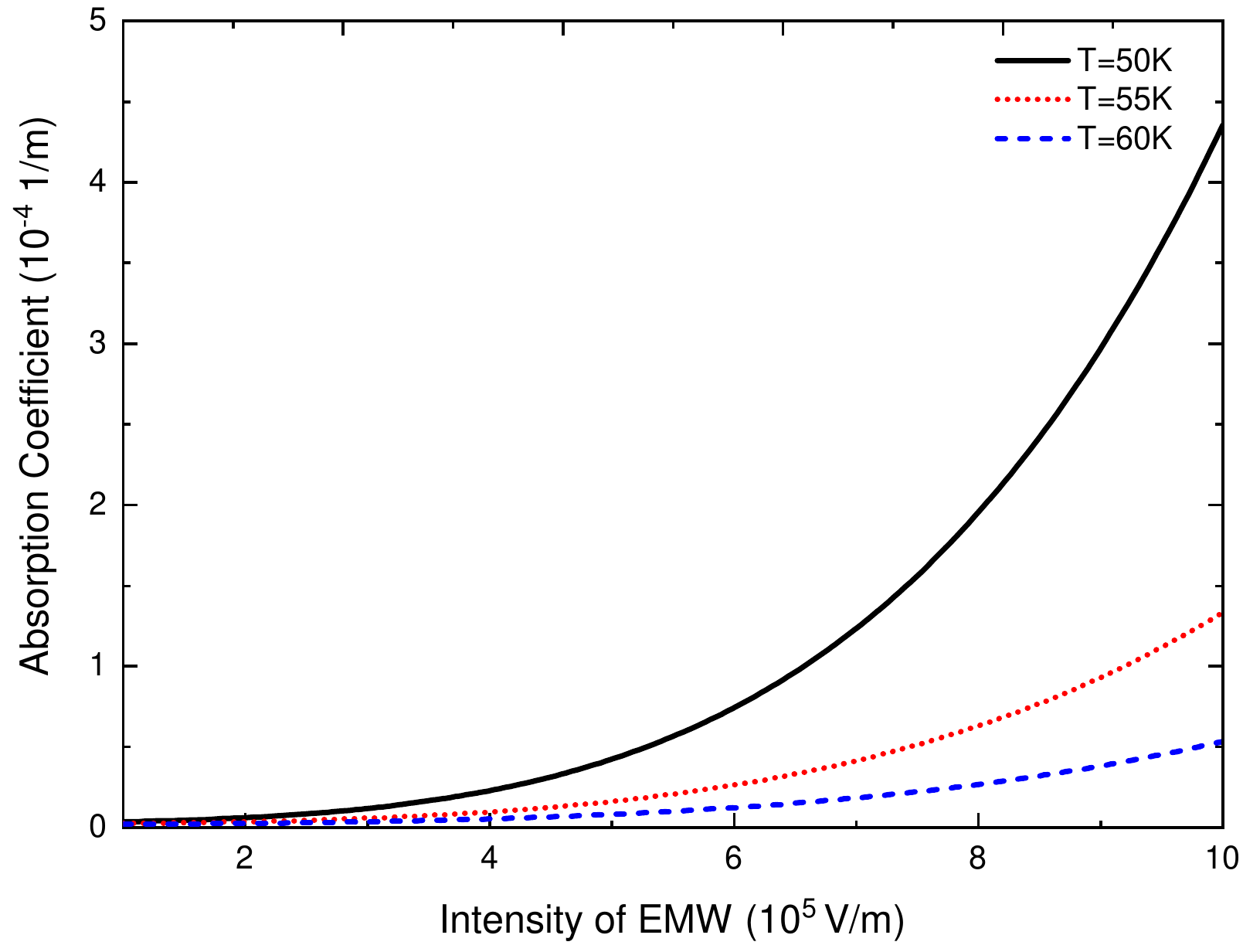}}
\caption{(Color online) The dependence of $\alpha$ on the intensity of electromagnetic wave}
\label{ae0B}
\end{figure}

The MNAC is displayed for both interactions of phonons as the function of the intensity of EMW  in Fig. \ref{ae0B}. We can quickly conclude that the absorption coefficient gradually rises where $E_0$ increases. Different from normal bulk semiconductors \cite{eps, eps1}, Quantum Wells \cite{bau} and Quantum Wires \cite{bau3}, the non-linear absorption coefficient in graphene increases stronger because electrons in graphene are relativistic Dirac-fermions with a nearly linear energy spectrum.

\begin{figure}
\centering
\subfigure[][Electron – Optical phonon interaction \label{aohw0b}]
  {\includegraphics[width=0.45\linewidth]{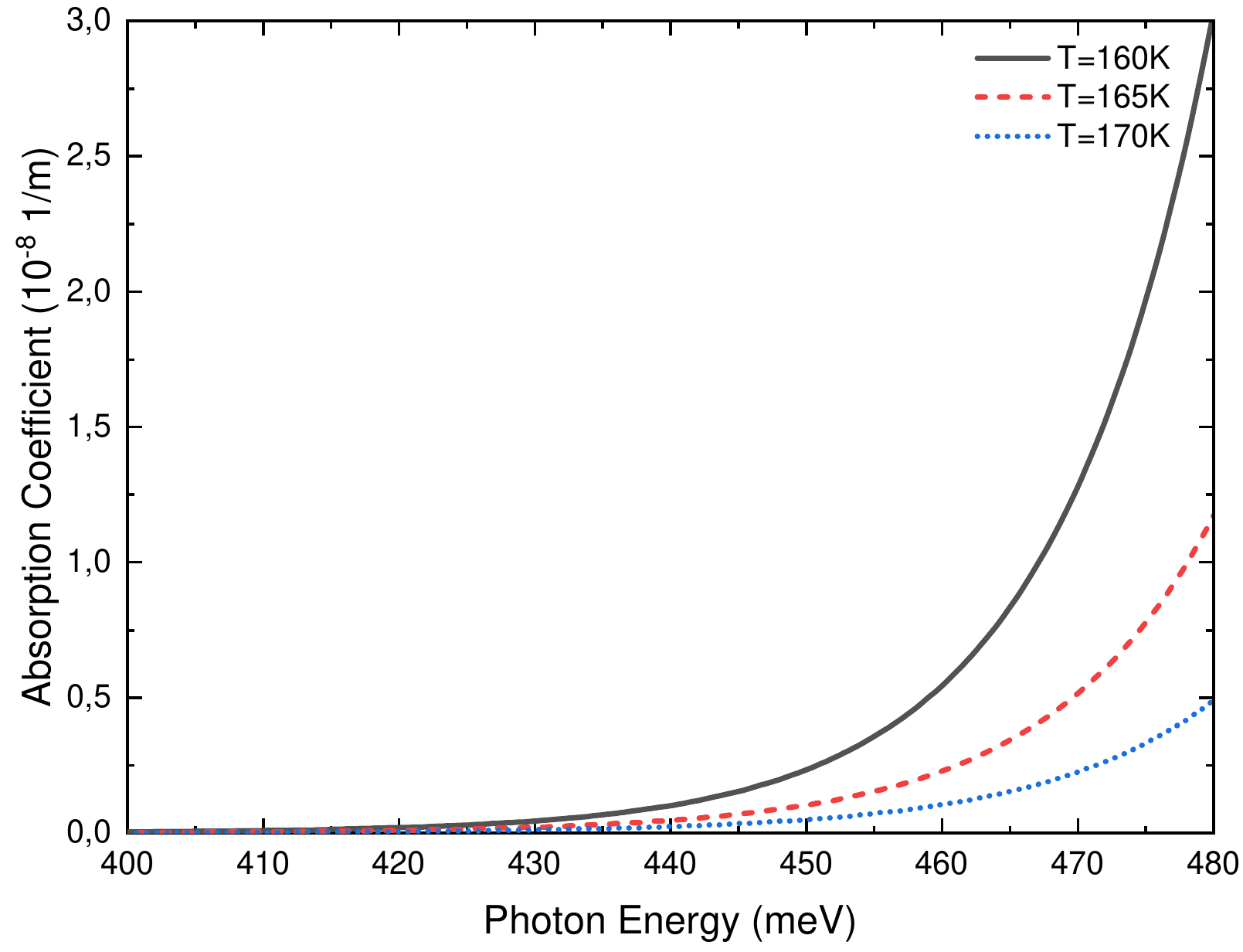}}
\subfigure[][Electron – Acoustic phonon interaction\label{aachw0b}]
  {\includegraphics[width=0.45\linewidth]{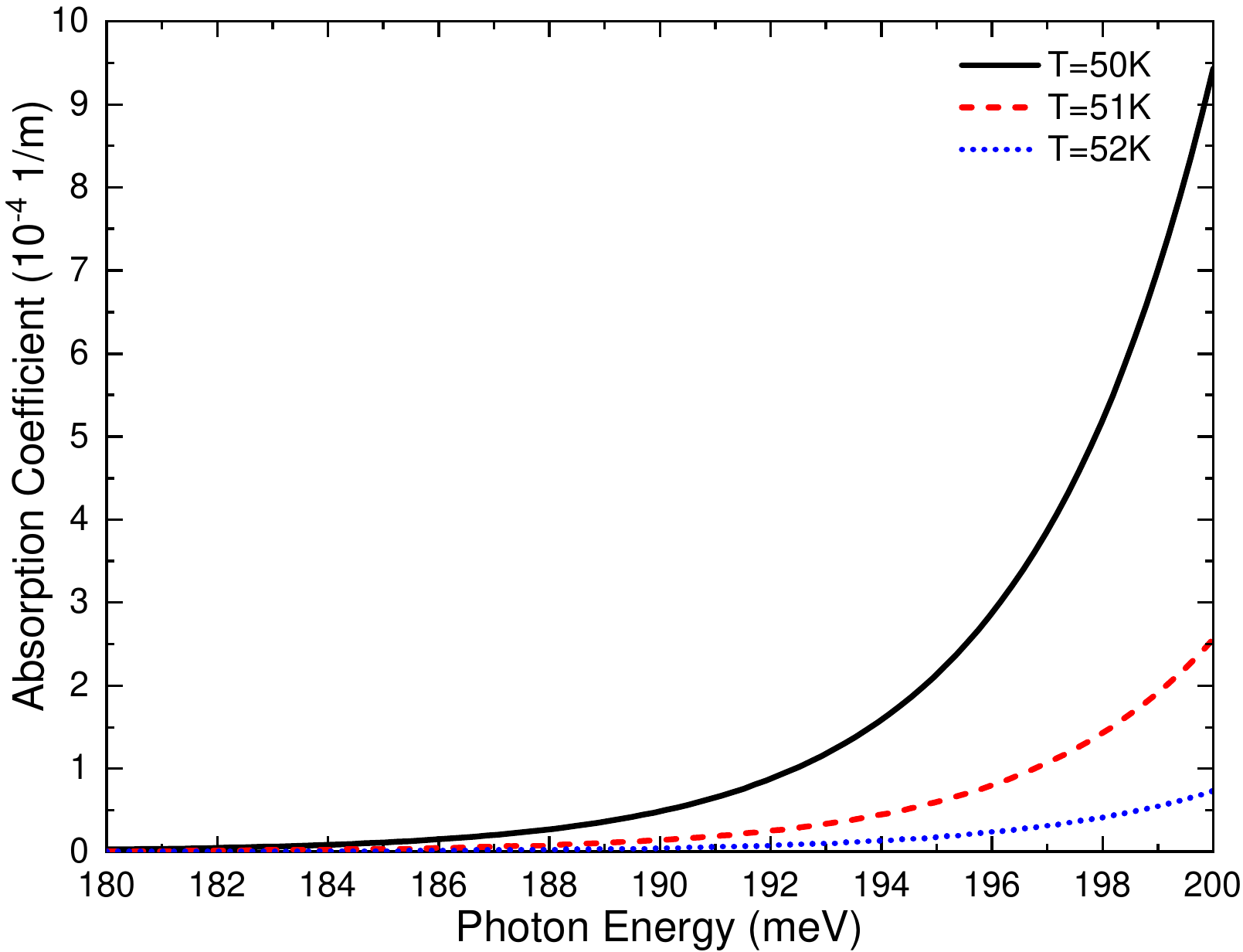}}
\caption{(Color online) The dependence of $\alpha$ on the photon energy of electromagnetic wave. Here, ${E_0} = 5 \times {10^5}{\rm{V/m}}$}
\label{ahw0B}
\end{figure}

We investigate and graph the influence of the absorption coefficient on photon energy and three different temperature cases in Fig. \ref{ahw0B}. We can easily obtain that the absorption coefficient goes up gradually as the temperature rises. This differs from that for other 2D systems \cite{bau, phong} (for example, in Quantum Wells, the absorption coefficient reaches one maximum peak). This results from the difference between the electron structure of Graphene and Quantum Wells. Specifically, the energy spectrum of electrons in graphene is linearly dependent on the k-wave vector, while the electron energy spectrum in traditional semiconductor systems is parabolic. Moreover, in the case of electron-acoustic phonon interaction, when the temperature is greater than 50K, the absorption coefficient has a small value, which shows that the contribution of this interaction process in the high-temperature domain is negligible. 

\subsection{In the Presence of an External Magnetic Field}

\begin{figure}
\centering
\subfigure[][Electron – Optical phonon interaction \label{aoTB}]
  {\includegraphics[width=0.45\linewidth]{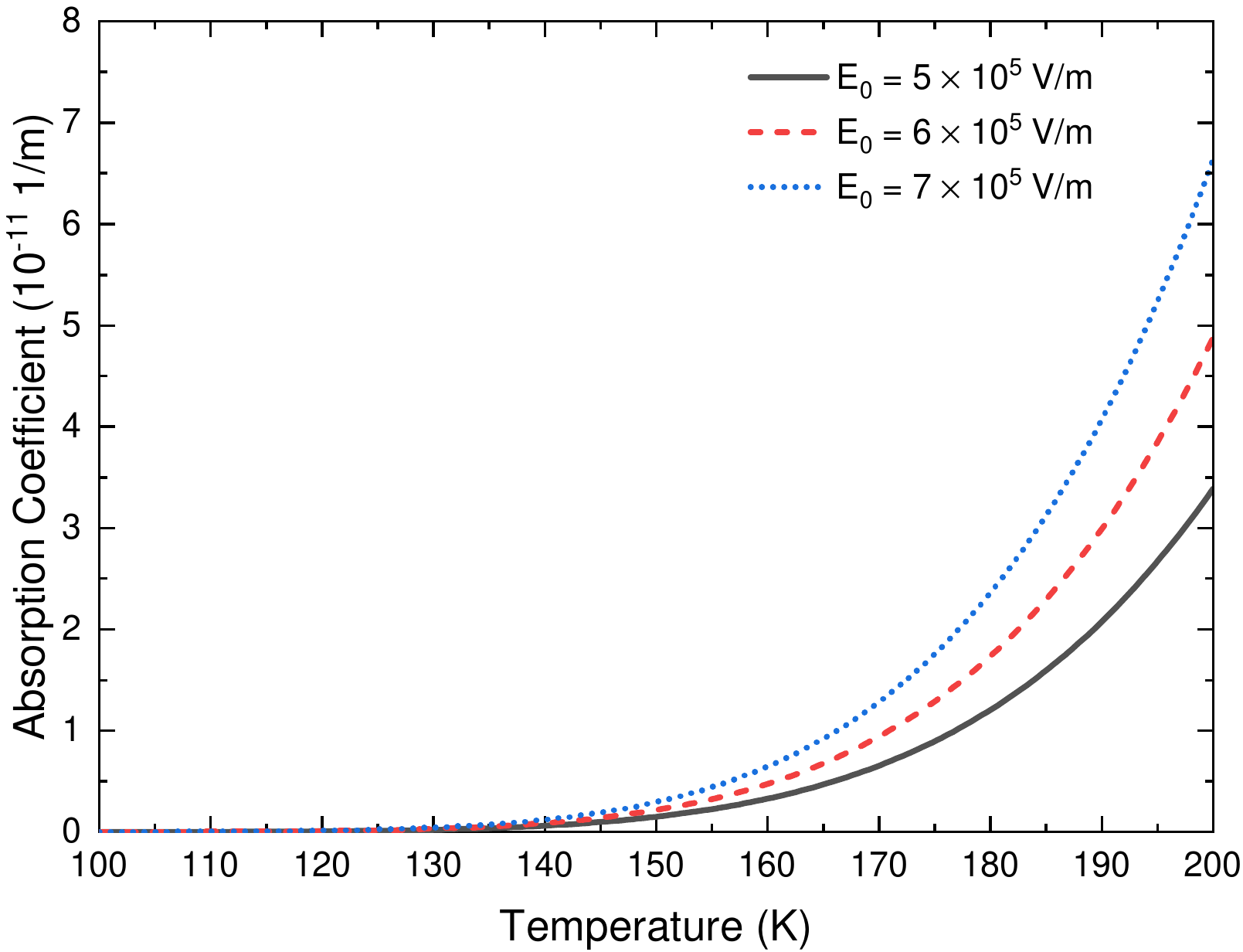}}
\subfigure[][Electron – Acoustic phonon interaction \label{aaTB}]
  {\includegraphics[width=0.45\linewidth]{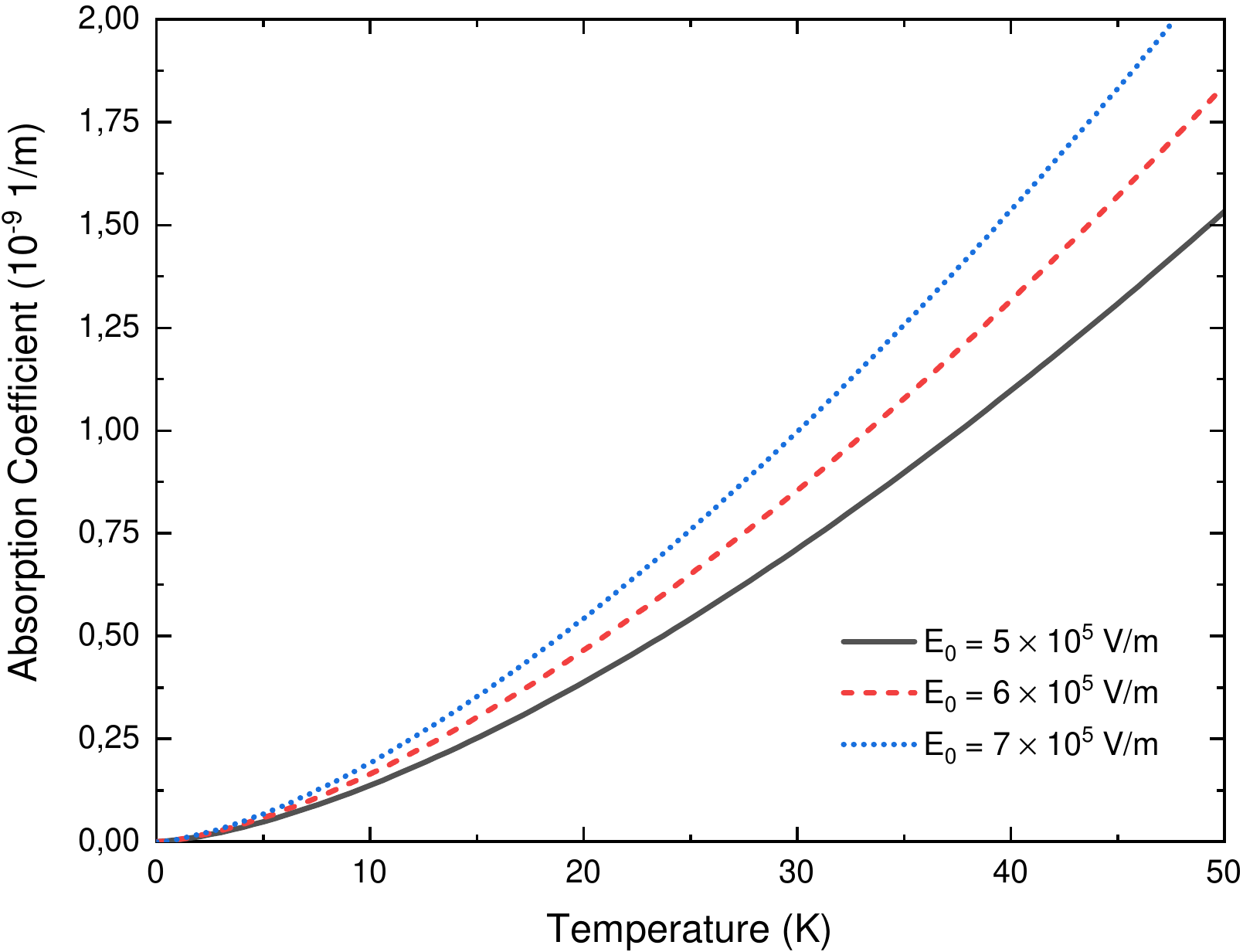}}
\caption{(Color online) The dependence of $\alpha$ on the temperature of system. Here, $\hbar \Omega  = 250{\rm{meV}}, B=6 {\rm{T}}$}
\label{aTB}
\end{figure}

In Fig. \ref{aTB}, we show the dependence of the MNAC on the temperature $T$ with two scattering mechanisms at different values of the intensity of EMW $E_0$. In all two cases (Fig. \ref{aoTB} and Fig. \ref{aaTB}), we realize that this dependence is non-linear, as we mentioned above. As the intensity of EMW increases, the MNAC also increases. For the case of electron-optical phonon interaction (Fig. \ref{aoTB}), we can see that the MNAC depends almost linearly on temperature. This is consistent with the result obtained in the case of a Quantum Well (another two-dimensional system) \cite{bau,bau3}.

\begin{figure}
\centering
\subfigure[][Consider 2 different temperature values: $B=6 {\rm{T}}$ \label{aohwTB}]
  {\includegraphics[width=0.45\linewidth]{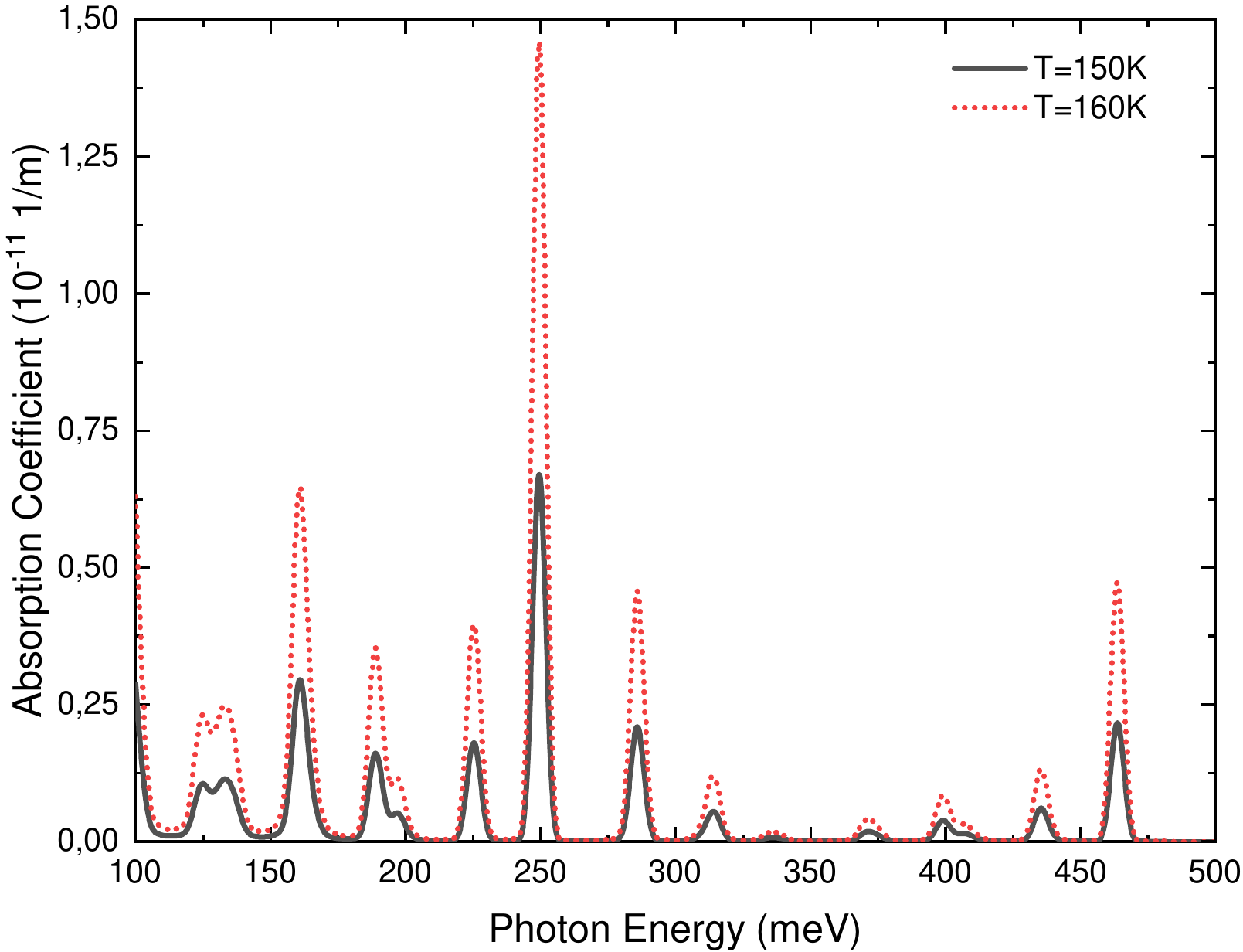}}
\subfigure[][Consider 2 different magnetic field values: $T=100 {\rm{K}}$  \label{aOhwB}]
  {\includegraphics[width=0.45\linewidth]{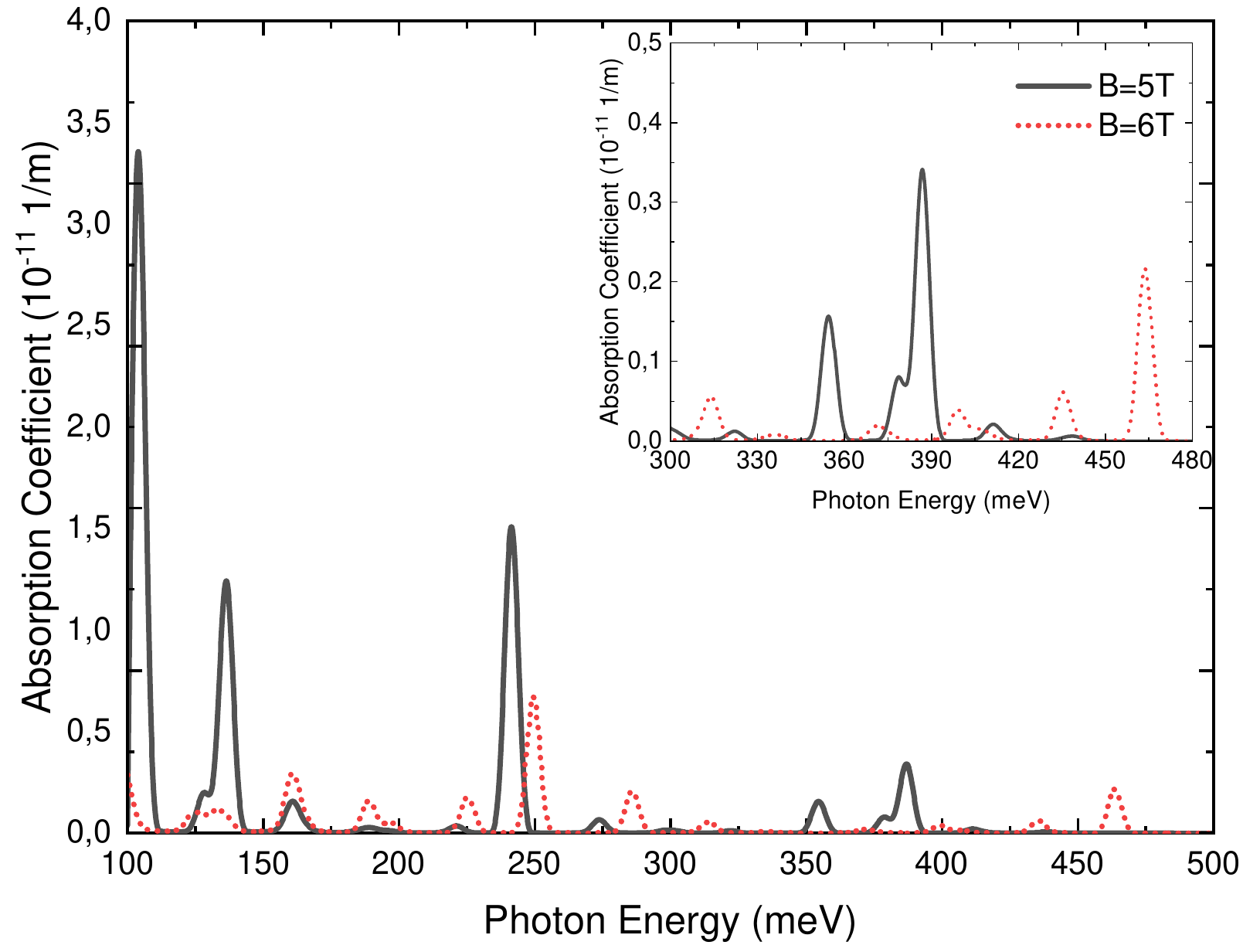}}
\caption{(Color online) The dependence of $\alpha$ on the photon energy of electromagnetic wave (electron – optical phonon interaction). Here, $E_{0}=5 \times 10^{5} \rm{V/m}$}
\label{aohw}
\end{figure}

\begin{figure}
\centering
\subfigure[][Consider 2 different temperature values: $B=6 {\rm{T}}$ \label{aachwTB}]
  {\includegraphics[width=0.45\linewidth]{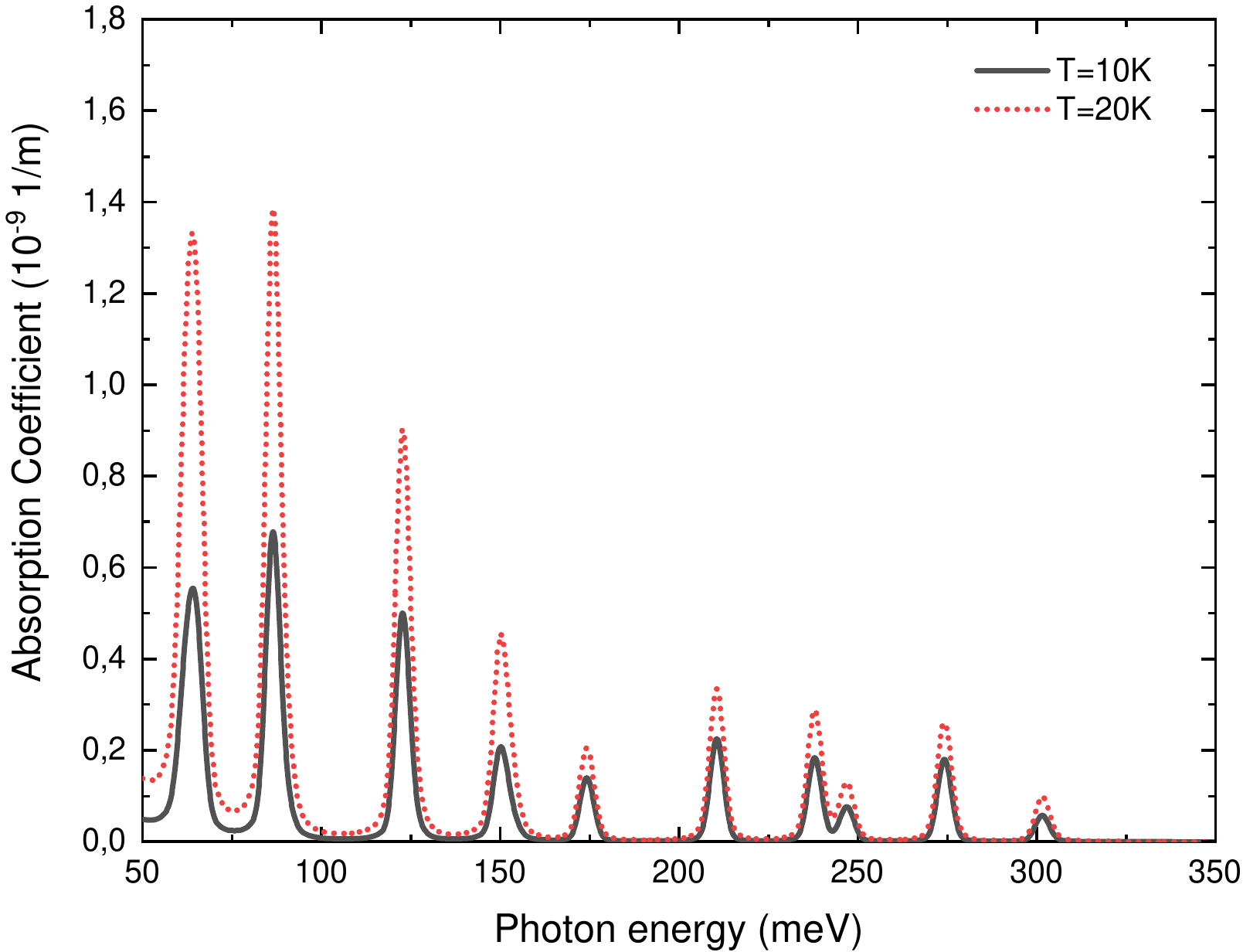}}
\subfigure[][Consider 2 different magnetic field values: $T=20 {\rm{K}}$  \label{aachwB}]
  {\includegraphics[width=0.45\linewidth]{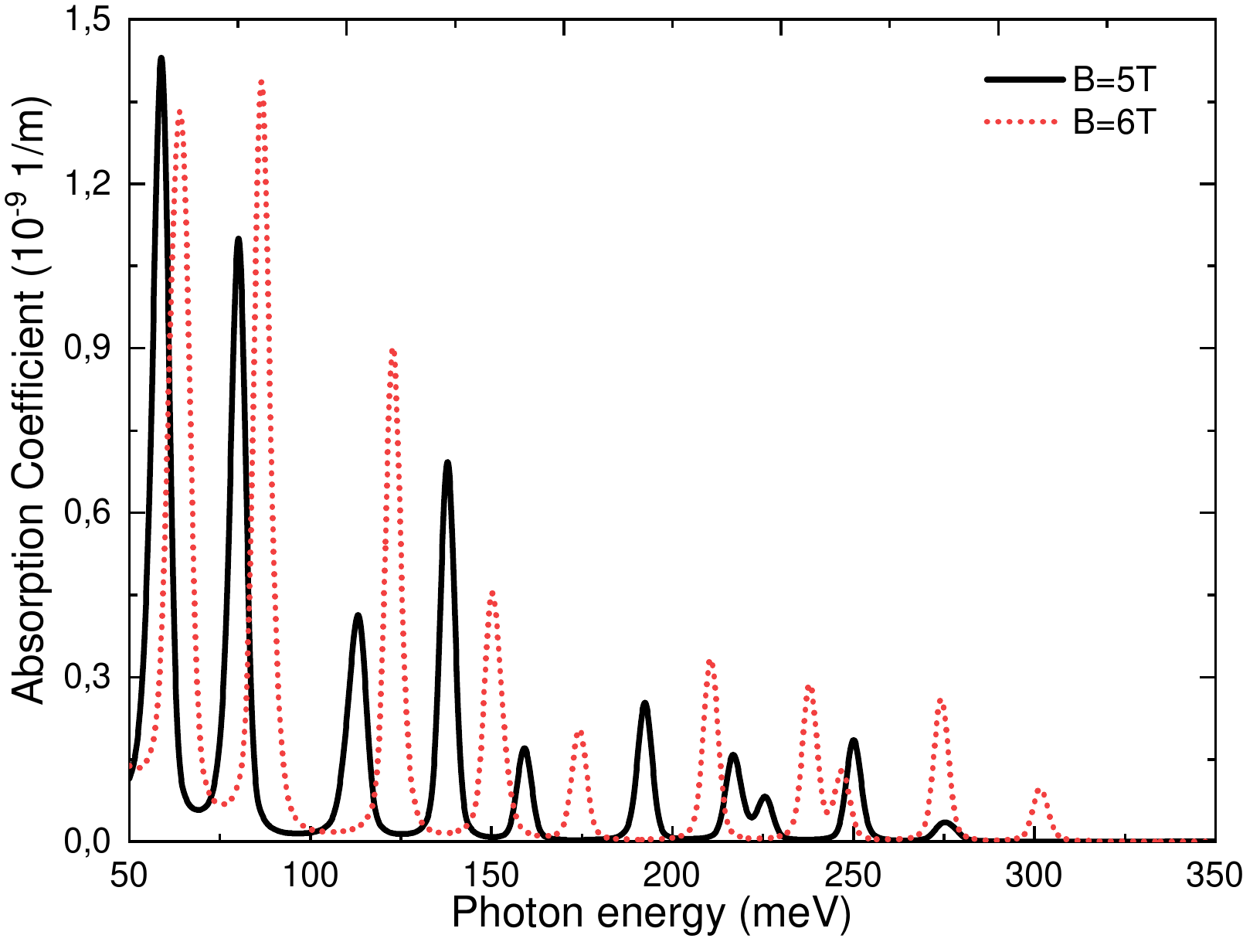}}
\caption{(Color online) The dependence of $\alpha$ on the photon energy of electromagnetic wave (electron – acoustic phonon interaction). Here, $E_{0}=5 \times 10^{5} \rm{V/m}$}
\label{aachw}
\end{figure}

In Fig. \ref{aohw} and Fig. \ref{aachw}, the non-linear absorption coefficient is plotted versus the photon energy $\hbar \Omega$ with two scattering mechanisms at two different values of the temperature. From Fig. \ref{aohwTB} and Fig. \ref{aachwTB}, we can see the appearance of absorption peaks related to the state transition of the electrons when absorbing photons of the electromagnetic wave in the electron-phonon interaction picture. The important thing in these two figures is that the position of the peaks does not depend on the temperature of the system but only shifts to the left (the position with the higher photon energy value) when the magnetic field strength is increased (this is shown on figures \ref{aOhwB} and \ref{aachwB}). In other words, the temperature only affects the value of the peak but not its position; the higher the temperature, the higher the absorption peaks. The position of the resonance peaks in the electron-optical phonon scattering obeys the cyclotron–phonon–photon resonance condition $\hbar {\omega _B}\left( {{S_{n'}}\sqrt {\left| {n'} \right|}  - {S_n}\sqrt {\left| n \right|} } \right) + \hbar {\omega _0} - \ell \hbar \Omega  = 0$ while in electron-acoustic phonon scattering they obey the cyclotron-phonon-photon resonance condition $\hbar {\omega _B}\left( {{S_{n'}}\sqrt {\left| {n'} \right|}  - {S_n}\sqrt {\left| n \right|} } \right) - \ell \hbar \Omega  = 0$ with $\ell  = 1,2,3$.  This result is similar to the results obtained in traditional semiconductor systems \cite{bau, bau3, phong, phong1}. In particular, the results of absorption of 1 and 2 photons (peak (1) and peak (7), (8)) related to the principal transitions in electron-optical phonon electron scattering under the effect of the magnetic field $B = 6 \rm{T}$ in Ref. \cite{phuc} were obtained here (see Fig. \ref{aohwTB}). From Fig. \ref{aohwTB}, we can also observe the 3-photon absorption peak (peak (2)) due to the electron's transition from $n = -1$ to $n' = 3$, satisfying ${\Delta _{3, - 1}} + \hbar {\omega _0} = 3\hbar \Omega$ with $\hbar \Omega  \approx 133{\rm{meV}}$.

\begin{figure}
\centering
\subfigure[][Electron – Optical phonon interaction: $T=100 {\rm{K}}$ \label{aohwB}]
  {\includegraphics[width=0.45\linewidth]{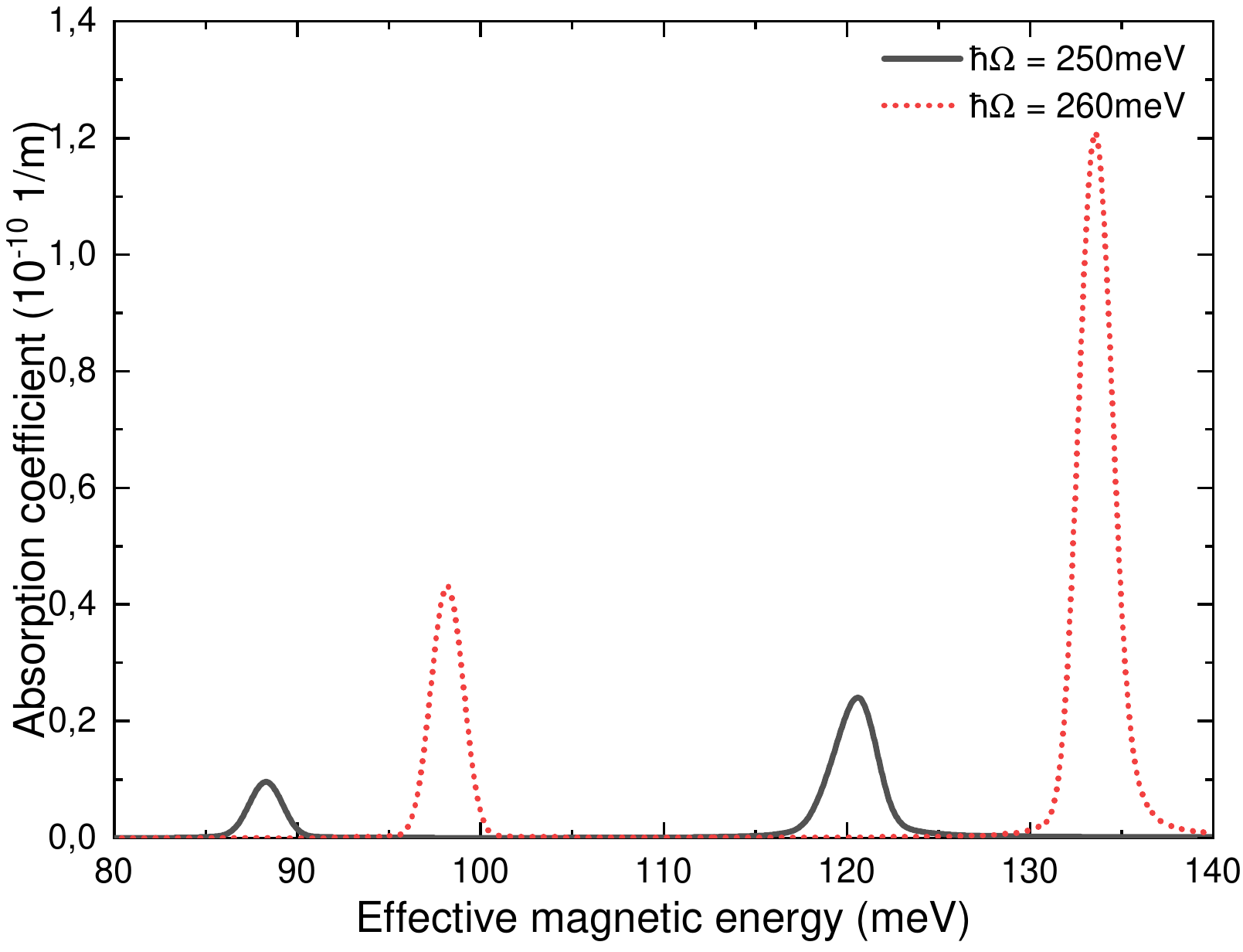}}
\subfigure[][Electron – Acoustic phonon interaction: $T=20 {\rm{K}}$ \label{aahwB}]
  {\includegraphics[width=0.45\linewidth]{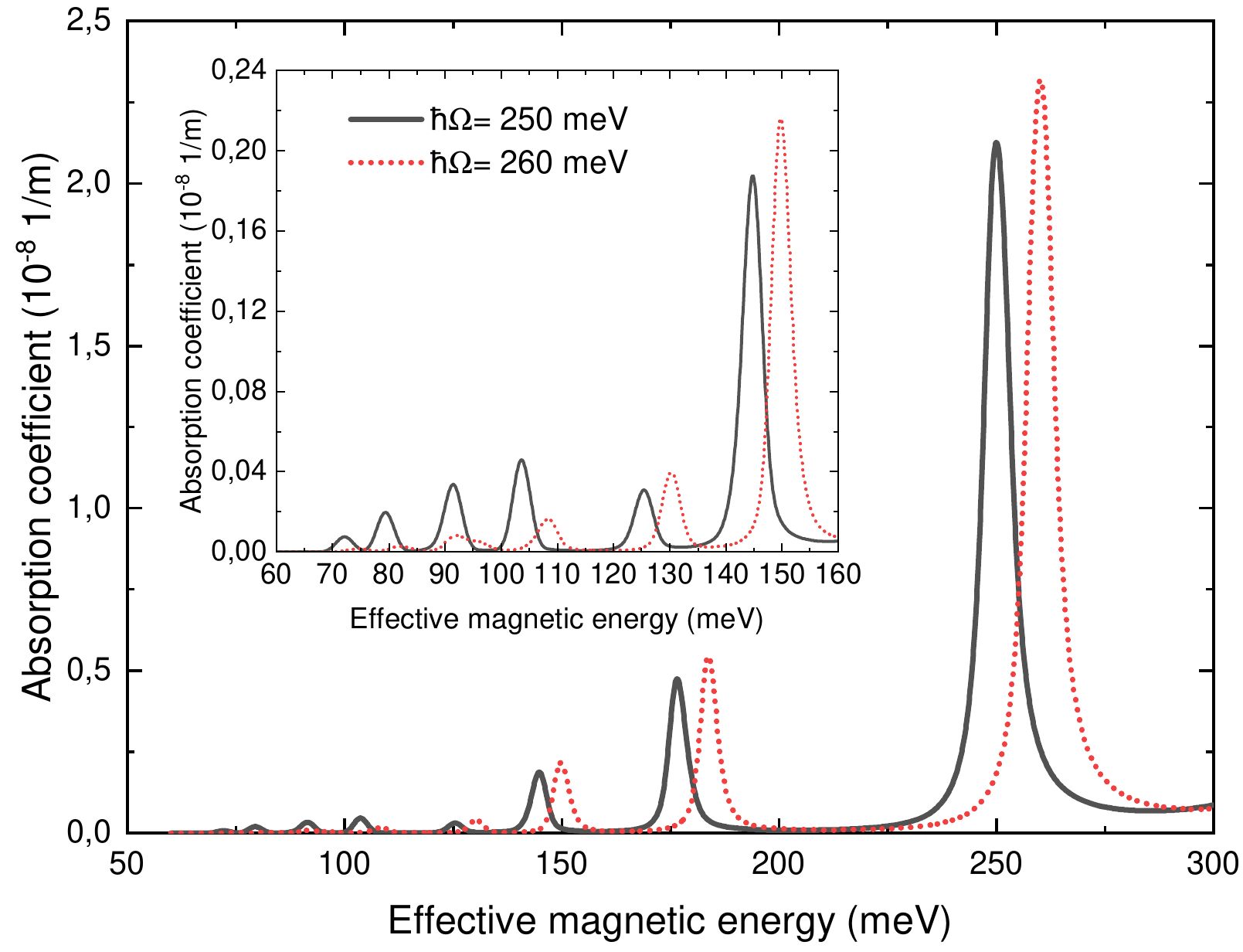}}
\caption{(Color online) The dependence of $\alpha$ on the magnetic energy. Here, $E_{0}=5 \times 10^{5} \rm{V/m}$}
\label{ahwB}
\end{figure}

Fig. \ref{ahwB} indicates the MNAC as a function of the effective magnetic energy (proportional to the square root of magnetic field B) at two different photon energy values. As seen from Fig. \ref{ahwB}, we see the appearance of spectral absorption lines. The absorption coefficient is only significant at these resonance peak positions. As mentioned above, the appearance of resonance peaks can be explained by cyclotron-phonon resonance conditions. In addition, the density of absorption peaks becomes sparse as the effective magnetic energy increases. This can be explained by the influence of the strong magnetic field causing the absorption spectrum to be interrupted in the region where the effective magnetic energy is greater than the photon energy of the electromagnetic wave. 

\begin{figure}
\centering
\includegraphics[width=10.0cm]{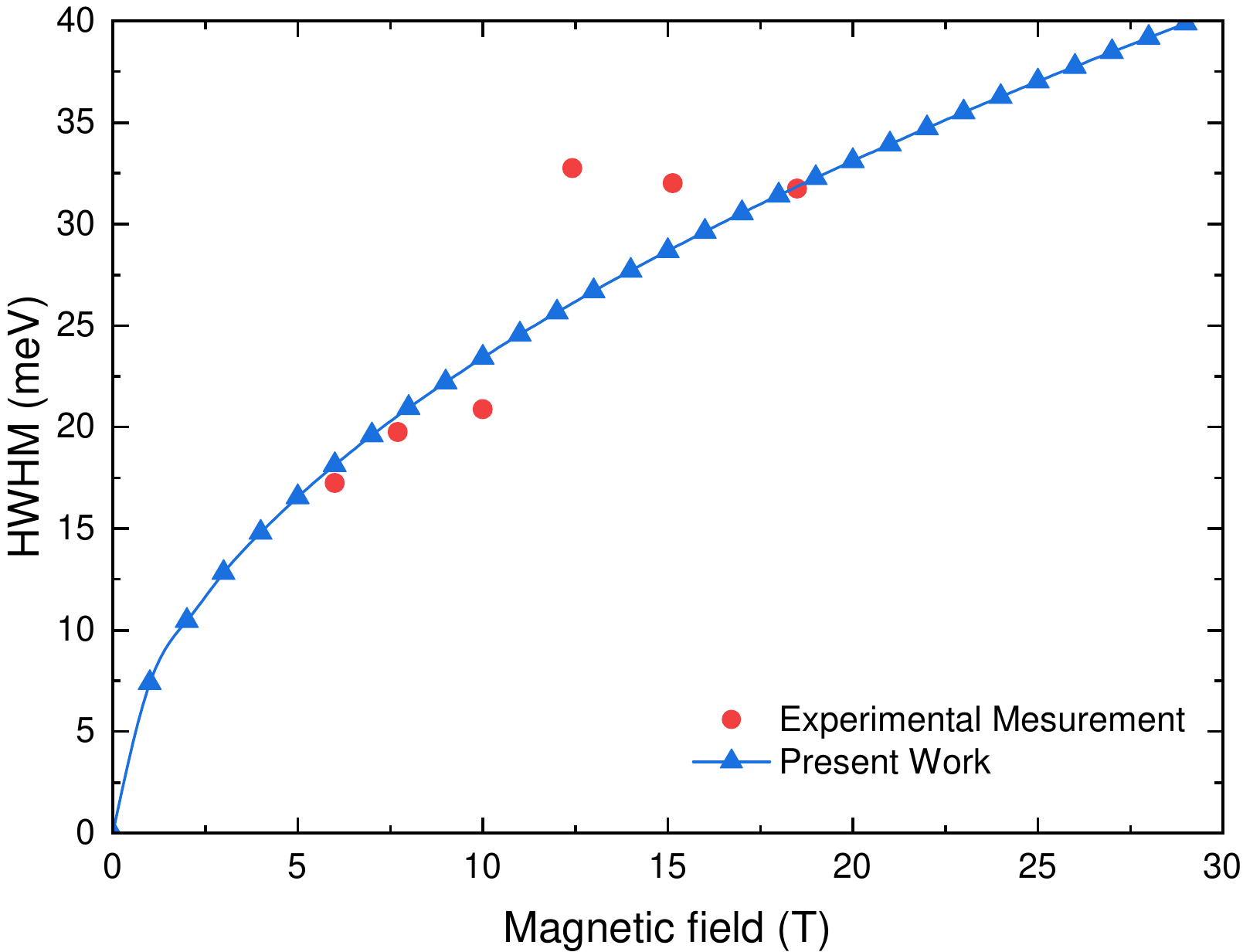}
\caption{(Color online) The dependence of HWHM on the magnetic field at T = 4.2K for the transition $n=0$ and $n'=1$. The squares and circles are our calculation, and the experimental data taken from ref. \cite{ji}}\label{hwhm}
\end{figure}

We now investigate the dependence of the Half Width at Half Maximum (HWHM) on the magnetic field in Fig. \ref{hwhm}. In this figure, we show a comparison between the present results and the experimental results obtained previously by other authors \cite{ji}. The results on the graph show a relatively good agreement between our experimental measurements and our theoretical calculations. Theoretically, the dependence of the HWHM on the magnetic field also follows the square root law of the form ${\mathcal{HWHM}} \approx {\rm{7}}{\rm{.40}}\sqrt {{\rm{B(T)}}} {\rm{(meV)}}$, which is also consistent with the theoretical calculations based on the projection operator method \cite{h1} and the perturbation theory \cite{phuc}. 
\section{Conclusions}
To sum up, we have theoretically investigated the Quantum Multi-photon Non-linear Absorption Coefficient in monolayer graphene with two scattering mechanisms based on the Quantum kinetic equation method. The new theoretical expressions for the absorption coefficient in graphene are established as functions of the external field, the photon energy of EMW, the system's temperature, and the magnetic field. The phenomenon of electromagnetic wave absorption in graphene has been studied in detail from the low-temperature domain (the limitation of classical theories) to the high-temperature domain, giving results in good agreement with classical theories such as the Boltzmann kinetic equation. 

In the presence of a strong magnetic field, the absorption spectrum of electromagnetic waves is interrupted. Resonance peaks appear according to cyclotron-phonon resonance conditions. The position of the resonance peaks depends only on the magnetic field and not on the system's temperature. The contribution of processes that absorb more than one photon is shown more clearly than other methods. The influence of a strong magnetic field interrupts the absorption spectrum in the highly effective magnetic energy domain. In addition, the broadening of the absorption spectrum under the increasing influence of the external magnetic field obeys the square root law of the form ${\mathcal{HWHM}} \approx {\rm{7}}{\rm{.40}}\sqrt {{\rm{B(T)}}} {\rm{(meV)}}$, which is in good agreement with experimental observations and previous theoretical studies. This result can be easily used to check the fabrication accuracy of graphene-related electronic devices. 

\begin{acknowledgments}
This research is financial by Vietnam National University, Hanoi - Grant number QG.22.11. 
\end{acknowledgments}
\appendix*
\section{The integrals used in the calculation}
From Eqs. A1, A2, A3 and A4 of Ref. \cite{phuc}, we have 
\begin{align}
{I_1} &= \frac{{m!}}{{\left( {m + j} \right)!}}\int_0^{ + \infty } {{e^{ - u}}{u^{j + 3}}{{\left[ {L_m^j\left( u \right)} \right]}^2}du} \\
 &= \left( {2m + j + 3} \right)\left\{ {2 + 6m\left( {m + 1} \right) + j\left[ {j + 3\left( {2m + 1} \right)} \right]} \right\} + 4m\left( {2m + j} \right)\left( {m + j} \right) \nonumber \\
 {I_2} &= \frac{{\left( {m - 1} \right)!}}{{\left( {m + j} \right)!}}\int_0^{ + \infty } {{e^{ - u}}{u^{j + 3}}L_m^j\left( u \right)L_{m - 1}^j\left( u \right)du}  =  - 3\left( {1 + {j^2} + 5mj + 5{m^2}} \right)
\end{align}

\bibliography{refs.bib}
\newpage
%Fig. 1
%Fig. 2
%Fig. 3
%Fig. 4
%Fig. 5
%Fig. 6
%Fig. 7
%Fig. 8 
\end{document}